\documentclass[prd,aps,twocolumn,preprintnumbers, showpacs, nofootinbib,superscriptaddress,notitlepage]{revtex4-1}
\usepackage{mathrsfs}
\usepackage{amsfonts}
\usepackage{amsmath}
\usepackage{slashed}
\usepackage{array}
\usepackage{verbatim}
\usepackage{epsfig}
\usepackage{graphicx}
\usepackage{color}
\usepackage[dvipsnames]{xcolor}
\usepackage[colorlinks,linkcolor=red,anchorcolor=green,citecolor=blue,CJKbookmarks=True]{hyperref}
\definecolor{red}{rgb}{1,0,0}
\newcommand{\red}[1]{{\color{red} #1}}

\usepackage{multirow}

\newcommand{\beq}{\begin{eqnarray}}
\newcommand{\eeq}{\end{eqnarray}}

\def\be{\begin{equation}}
\def\ee{\end{equation}}
\def\bea{\begin{eqnarray}}
\def\eea{\end{eqnarray}}
\def\bal#1\eal{\begin{align}#1\end{align}}

\begin{document}

\title{Comparison of the mixed-fermion-action Effects using different fermion and gauge actions with 2+1 and 2+1+1 flavors}

\collaboration{\bf{CLQCD Collaboration}}

\author{
\includegraphics[scale=0.30]{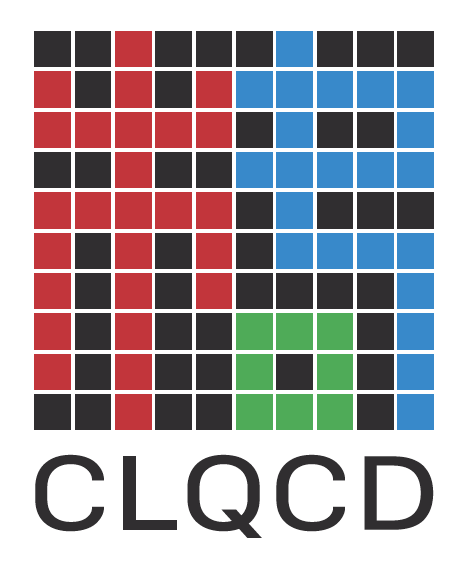}\\
Zun-Xian Zhang}
\affiliation{School of Fundamental Physics and Mathematical Sciences, Hangzhou Institute for Advanced Study, UCAS, Hangzhou 310024, China}
\affiliation{CAS Key Laboratory of Theoretical Physics, Institute of Theoretical Physics, Chinese Academy of Sciences, Beijing 100190, China}
\affiliation{University of Chinese Academy of Sciences, Beijing 100049, China}

\author{Mengchu Cai}
\affiliation{CAS Key Laboratory of Theoretical Physics, Institute of Theoretical Physics, Chinese Academy of Sciences, Beijing 100190, China}

\author{Bolun Hu}
\affiliation{Computation-based Science and Technology Research Center, The Cyprus Institute, 20 Kavafi Str., Nicosia 2121, Cyprus}

\author{Xiangyu Jiang}
\affiliation{Department of Physics, Indiana University, Bloomington, Indiana 47405, USA}

\author{Xiao-Lan Meng}
\affiliation{CAS Key Laboratory of Theoretical Physics, Institute of Theoretical Physics, Chinese Academy of Sciences, Beijing 100190, China}
\affiliation{University of Chinese Academy of Sciences, School of Physical Sciences, Beijing 100049, China}

\author{Yi-Bo Yang}
\email[Corresponding author: ]{ybyang@itp.ac.cn}
\affiliation{CAS Key Laboratory of Theoretical Physics, Institute of Theoretical Physics, Chinese Academy of Sciences, Beijing 100190, China}
\affiliation{University of Chinese Academy of Sciences, School of Physical Sciences, Beijing 100049, China}
\affiliation{School of Fundamental Physics and Mathematical Sciences, Hangzhou Institute for Advanced Study, UCAS, Hangzhou 310024, China}
\affiliation{International Centre for Theoretical Physics Asia-Pacific, Beijing/Hangzhou, China}

\author{Dian-Jun Zhao}
\email[Corresponding author: ]{zhaodianjun@cuhk.edu.cn}
\affiliation{School of Science and Engineering, The Chinese University of Hong Kong, Shenzhen 518172, China}

\date{\today}

\begin{abstract}
The leading-order low-energy constant $\Delta_{\rm mix}$ in mixed-action chiral perturbation theory is calculated using $2+1+1$-flavor gauge ensembles with HISQ fermions and a tadpole-improved Symanzik gauge action at four lattice spacings $a \in [0.048, 0.111]$ fm. By comparing our results to those from different actions and a $2+1$-flavor case, We find that the fermion action has the dominant impact, the gauge action has a secondary but measurable effect, and the contribution from charm quark loops is negligible within our current uncertainties.
\end{abstract}

\maketitle

\section{Introduction}\label{sec:intro}

Lattice QCD provides the primary non-perturbative framework for \textit{ab initio} calculations of strong interaction physics. A central challenge in the field is the choice of fermion discretization, which must balance computational cost against the preservation of fundamental symmetries, most notably chiral symmetry. In this context, the \textit{mixed-action} approach has emerged as a powerful pragmatic strategy. By generating gauge ensembles with a computationally efficient ``sea" fermion action and calculating observables using a more theoretically clean but expensive ``valence" fermion action, one can leverage the statistical advantages of large ensembles while maintaining good control over systematic errors. Mixed-action lattice QCD has been employed in calculations of glue helicity~\cite{Yang:2017}, nucleon charges~(e.g.,\cite{Chang:2018,Jang:2020,Park:2025nucleon}), hadron vacuum polarization~\cite{Borsanyi:2021}, and other observables.

The success of this approach hinges on a quantitative understanding of the additional discretization artifacts introduced by the action mismatch, known as \textit{mixed-action effects}. A key calculable quantity for quantifying these effects is the mass $m_{\pi,{\rm vs}}$ of the mixing pion with one valence quark and one sea anti-quark, which leads to the leading-order low-energy constant in the mixed-action partially-quenched chiral perturbation theory (MAPQ$\chi$PT)~\cite{Bar:2004,Bar:2005tu,Chen:2007ug}:
\begin{equation}
\label{eq:mix_act}
\Delta_{\text{mix}} \equiv m_{\pi,\text{vs}}^{2} - \frac{m_{\pi,\text{vv}}^{2} + m_{\pi,\text{ss}}^{2}}{2}|_{m_{\pi,\text{vv}}=m_{\pi,\text{ss}}},
\end{equation}
where $m_{\pi,\text{vv}}$ ($m_{\pi,\text{ss}}$) is the pion mass with the valence (sea) quark and anti-quark. 

In the past two decades, numerical lattice QCD studies of mixed-action effects have remained relatively sparse and fragmented~\cite{Durr:2007ef,Orginos:2007tw,Aubin:2008wk,Walker-Loud:2008rui,xQCD:2010pnl,Lujan:2012wg,Cichy:2012vg,Basak:2014kma,Berkowitz:2017opd,Basak:2017oup}. This is largely because such effects are typically treated as a component of the systematic uncertainty for specific valence-sea fermion combinations rather than as a primary subject of study. A systematic numerical investigation into the impact of specific valence and sea actions was absent until Ref.~\cite{Zhao:2022ooq}.

Based on high-precision calculations of $\Delta_{\text{mix}}$, Ref.~\cite{Zhao:2022ooq} established that when the sea fermion action preserves chiral symmetry, $\Delta_{\text{mix}}$ scales as $\mathcal{O}(a^4)$—a significant improvement over the expected $\mathcal{O}(a^2)$ behavior. This favorable scaling has been observed over a wide range of lattice spacings, $a \in [0.04, 0.19]$ fm, for chirally symmetric actions such as DW~\cite{Brower:2004xi,Brower:2005qw,Brower:2012vk} and HISQ fermions~\cite{Follana:2006rc}. In contrast, when a clover sea fermion action with explicit chiral symmetry breaking is used, $\Delta_{\text{mix}}$ remains substantially larger, independent of the valence action.

However, the comparison in Ref.~\cite{Zhao:2022ooq} presents potential loopholes. The clover fermion ensembles used the tadpole-improved tree-level Symanzik (TITLS) gauge action, while the HISQ and DW ensembles used different gauge actions (one-loop improved Symanzik and Iwasaki, respectively). Consequently, the possibility that the gauge action itself significantly impacts $\Delta_{\text{mix}}$ cannot be excluded. Furthermore, the number of dynamical fermion flavors differs between the clover ($N_f=2+1$) and HISQ ($N_f=2+1+1$) ensembles, introducing another confounding factor.

In this work, we perform a controlled study to isolate these effects. We generate new HISQ fermion ensembles using the same tadpole-improved Symanzik gauge action as the clover studies, with $N_f=2+1+1$ flavors at four lattice spacings, and three $N_f=2+1$ flavor ensembles at $a \sim 0.11$ fm with different gauge actions. By calculating $\Delta_{\text{mix}}$ on these new ensembles, we enable a fair comparison to precisely determine the impact of the gauge action and the dependence on the number of dynamical fermion flavors.

\section{Setup and basic tests of the generated gauge ensembles}\label{sec:setup}

The fermion action used for the gauge ensembles generated in this work is the HISQ action~\cite{Follana:2006rc}, which effectively mitigates taste symmetry breaking and discretization errors while retaining acceptable chiral symmetry. It is formulated in the following form:
\begin{equation}
S_{\text{HISQ}} = \sum_x \bar{\psi}(x)\left( \gamma \cdot D^{\text{HISQ}} + m \right)\psi(x),
\end{equation}
with the ${\cal O}(a^2)$ improved covariant derivative operator:
\begin{equation}
D^{\text{HISQ}}_\mu \equiv \Delta_\mu(W) - \frac{a^2}{6}(1 + \epsilon)\Delta_\mu^3(X),
\end{equation}
where the smeared link field $W_\mu(x)$ is constructed via a two-stage Fat7 smearing process with intermediate SU(3) projection, and $X_\mu(x)$ denotes links smeared with a single application of the ASQTAD smearing operator. The parameter $\epsilon$ provides improved dispersion relation for heavy quarks with bare mass $m$ by canceling tree-level $\mathcal{O}((am)^4)$ errors.

We consider a general gauge action of the form:
\begin{align}
S_g = \hat{\beta} \bigg[ & \sum_P \left( 1 - \frac{1}{3} \mathrm{Re}\mathrm{Tr}(P) \right) \nonumber\\
& + C_R \sum_R \left( 1 - \frac{1}{3} \mathrm{Re}\mathrm{Tr}(R) \right) \bigg],
\end{align}
where the sums run over all $1 \times 1$ plaquettes ($P$) and $2 \times 1$ rectangles ($R$). Specific gauge actions are characterized by the value of the parameter $C_R$:

\begin{itemize}
    \item $C_R = 0$ for the Wilson action (W);
    \item $C_R = -1/20$ for the tree-level Symanzik action ($S^{(0)}$);
    \item $C_R = -1/(20 u_0^2)$ for the tadpole-improved Symanzik action, with $u_0 = \left( \langle \mathrm{Re}\mathrm{Tr}(P) \rangle / 3 \right)^{1/4}$. This was employed in previous CLQCD ensembles with clover fermions~\cite{CLQCD:2023sdb,CLQCD:2024yyn};
    \item $C_R = -0.331/(1+8 \cdot 0.331) = -0.0907346$ for the Iwasaki gauge action.
\end{itemize}

\begin{table}[!h]
\centering
\caption{Input parameters $\hat{\beta}$, $u_0$, $\tilde{L}^3\times \tilde{T}$, and $\tilde{m}^{\rm b}_{l,s,c}$ of the 2+1+1 flavor gauge ensembles with HISQ fermion and $S^{\rm tad}$ gauge actions.}
\begin{tabular}{l|lcc|lll}
\hline
Ensemble & $\hat{\beta}$ & $\tilde{L}^3 \times \tilde{T}$ & $u_0$ &$\tilde{m}^{\rm b}_{l}$ & $\tilde{m}^{\rm b}_s$ &$\tilde{m}^{\rm b}_c$\\
\hline 
c24P31s &7.29&$24^3\times 48$&0.879440(3)&0.00944& 0.055 & 0.5555 \\
c24P31 &7.29&$24^3\times 48$&0.879452(3)&0.00944& 0.04721 & 0.5555 \\
c32P31 &7.29&$32^3\times 48$&0.879451(2)&0.00944& 0.04721 & 0.5555\\
c24P22 &7.29&$24^3\times 48$&0.879469(3)&0.00472& 0.04721 & 0.5555 \\
c32P22 &7.29&$32^3\times 48$&0.879468(2)&0.00472& 0.04721 & 0.5555 \\
c48P13 &7.29&$48^3\times 48$&0.879472(1)&0.00174& 0.04721 & 0.5555\\
e32P31 &7.54&$32^3\times 64$&0.886360(2)&0.007434& 0.03715 & 0.4371\\
g32P31 &7.75&$32^3\times 64$&0.891434(1)&0.00579& 0.02895 & 0.34 \\
g48P31 &7.75&$48^3\times 64$&0.891432(1)&0.00579& 0.02895 & 0.34\\
h48P31 &8.20 &$48^3\times 96$&0.900600(1)&0.003526& 0.01763 & 0.207\\
\hline
\end{tabular}
\label{tab:hisq_inputs}
\end{table}

In this work, we generated a set of 2+1+1-flavor gauge ensembles using the aforementioned fermion and gauge actions. The ensemble parameters are detailed in Table~\ref{tab:hisq_inputs} (with the symbol $\tilde{X}\equiv Xa^n$ for any quantity $X$ with mass dimension $n$), including the gauge coupling $\hat{\beta}$, the tadpole factor $u_0$, lattice volume $\tilde{L}^3\times \tilde{T}$, and the bare dimensionless quark masses $\tilde{m}^{\rm b}_{l,s,c}$. For all ensembles at $\hat{\beta}=7.29$ and 7.54 (except c24P31s to be discussed later), we fixed the charm-to-strange mass ratio to $\tilde{m}^{\rm b}_{c}/\tilde{m}^{\rm b}_{s}=11.766$, corresponding to the 2+1+1 flavor FLAG average~\cite{FlavourLatticeAveragingGroupFLAG:2024oxs,EuropeanTwistedMass:2014osg,Chakraborty:2014aca,FermilabLattice:2018est,ExtendedTwistedMass:2021gbo}. For the ensembles at $\beta=7.75$ and 8.20, we used $\tilde{m}^{\rm b}_{c}/\tilde{m}^{\rm b}_{s}=11.74$, a value consistent with the FLAG average of 11.766(30) within its uncertainty. The strange-to-light mass ratio, $\tilde{m}^{\rm b}_{s}/\tilde{m}^{\rm b}_{l}$, was set to 27.18~\cite{Bazavov:2017lyh} for the physical point ensemble (c48P13), to 10 for c24P22 and c32P22, and to 5 for the remaining ensembles (excluding c24P31s). The bare strange quark mass $\tilde{m}^{\rm b}_{s}$ was then tuned to ensure the resulting $m_{\eta_s}$ mass was close to the value of $689.63(18)$ MeV~\cite{Borsanyi:2020mff} which corresponds to the physical strange quark mass, and the tadpole factor $u_0$ was tuned to ensure consistency within 0.001\% between its input value and the value from thermalized configurations. Finally, to investigate the impact of the strange sea quark mass, we generated the c24P31s ensemble using the same $\hat{\beta}$ and $\tilde{m}^{\rm b}_{l,c}$ as the c24P31 ensemble but with a $\tilde{m}^{\rm b}_{s}$ enlarged by 17\%.

The values of $u_0$ in Table~\ref{tab:hisq_inputs} indicate a clear hierarchy in the parameter dependencies. The gauge coupling $\hat{\beta}$ is the dominant factor. Contributions from the light and strange quark masses, though much weaker (${\cal O}(10^{-5})$), are still relevant at our level of statistical uncertainty. Finally, any finite volume effect is negligible, remaining consistent with zero within a $2\sigma$ confidence interval.

\begin{figure}[hbt!]
\centering
\includegraphics[width=0.48\textwidth]{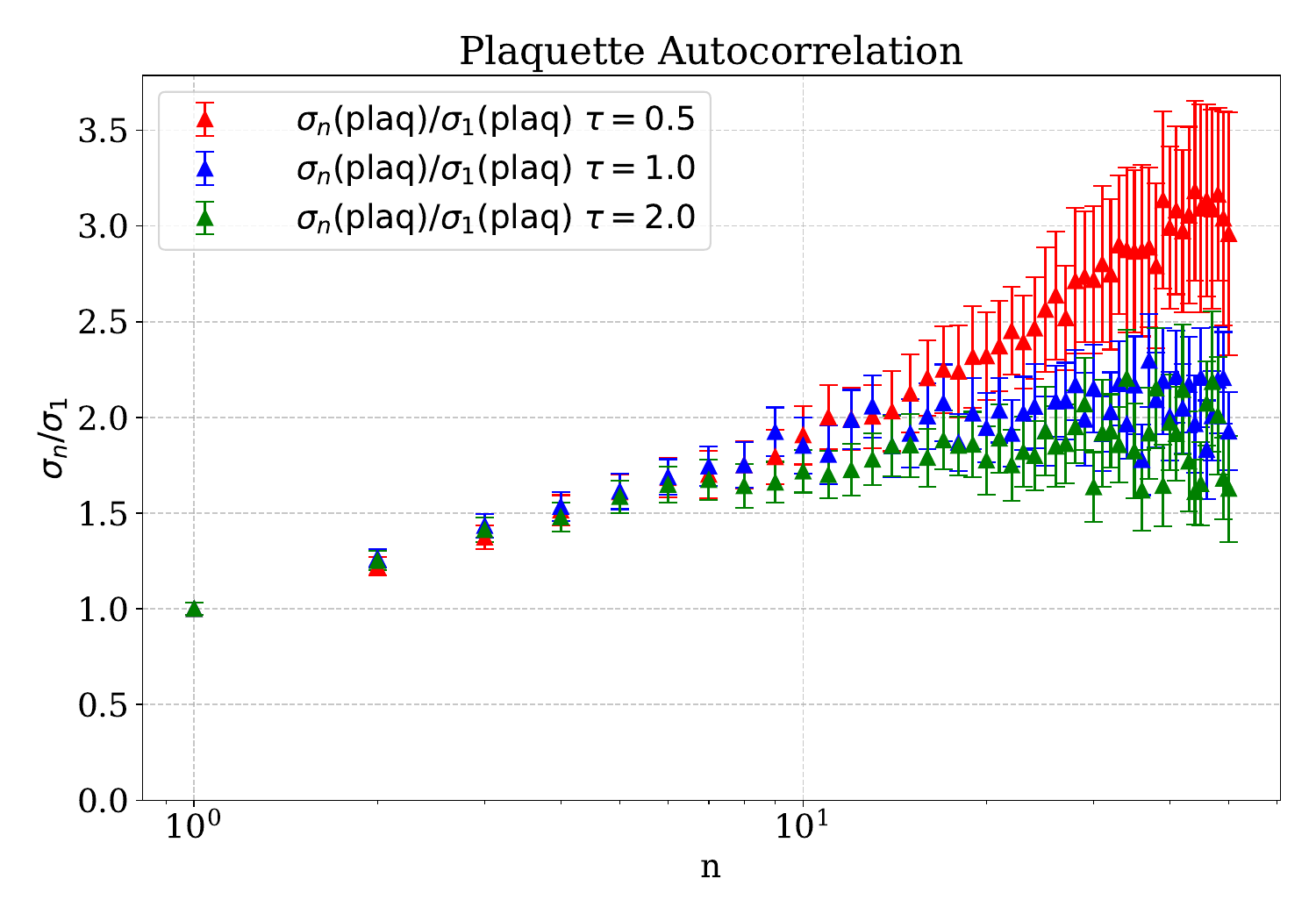}
\includegraphics[width=0.48\textwidth]{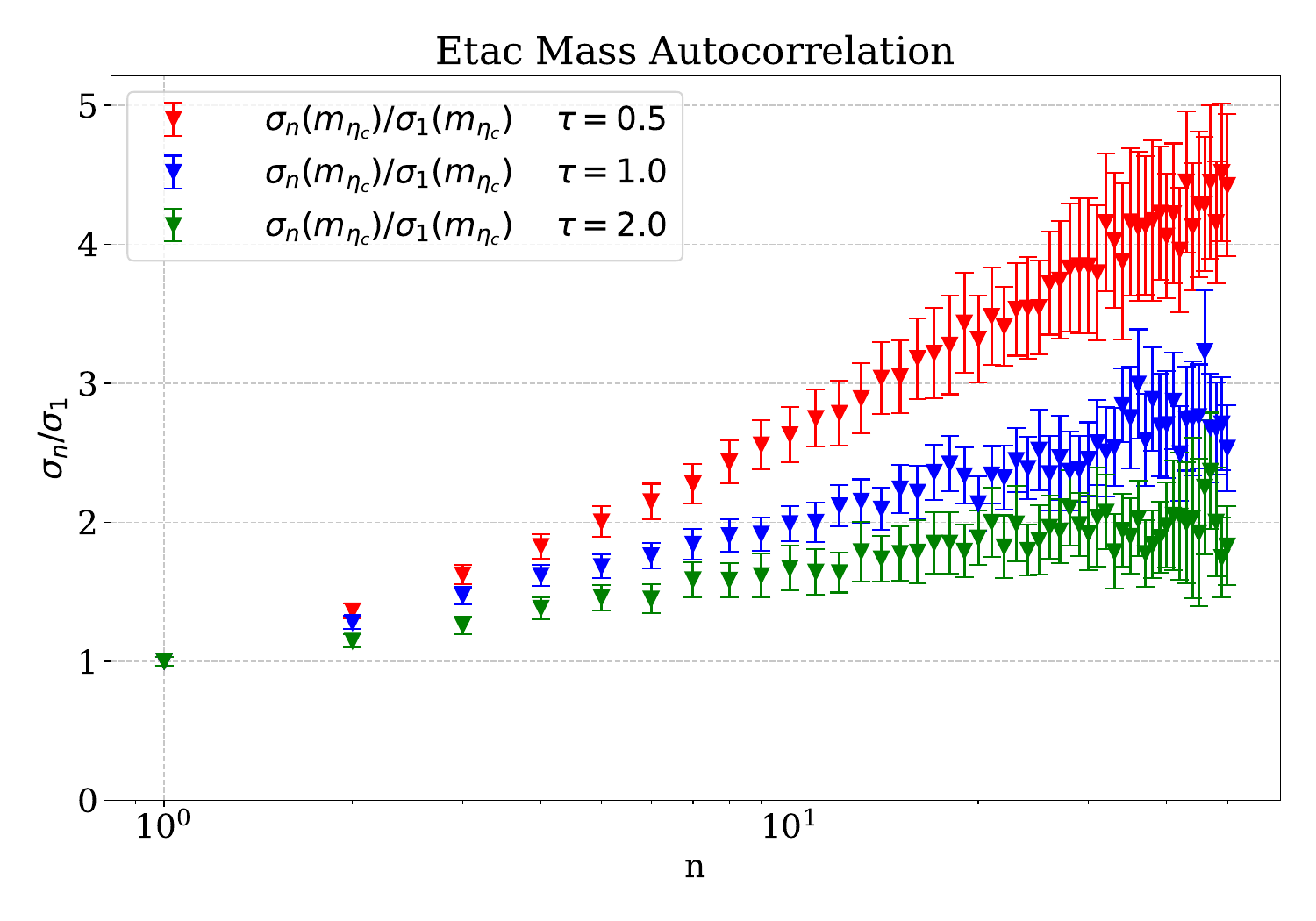}
\caption{The normalized variance ratio $\sigma_n/\sigma_1$ of the averaged plaquette $\mathrm{ReTr}(P)$ (upper panel) and $m_{\eta_c}$ (lower panel), using the molecular-dynamics time $\tau=0.5$ (red), 1.0 (blue) and 2.0 (green) per trajectory.}
\label{fig:autocorr1}
\end{figure}

To mitigate autocorrelation between configurations, we set the molecular dynamics time between trajectories to \(\tau = 2\) and employed multiple integration steps with \(\delta \tau = 0.02\) to maintain a high acceptance rate. We quantified autocorrelations by computing the normalized variance ratio \(\sigma_n/\sigma_1\) for various bin sizes \(n\), using the averaged plaquette \(\mathrm{ReTr}(P)\), the topological charge \(Q\), the pion mass \(m_\pi\), and the \(\eta_c\) mass \(m_{\eta_c}\), as detailed in the Appendix.

The results, \red{shown in Fig.~\ref{fig:autocorr1},} reveal that \(\mathrm{ReTr}(P)\) (upper panel) and \(m_{\eta_c}\) (lower panel) exhibit stronger autocorrelations, requiring larger bin sizes \(n\) for the variance ratio to saturate. Notably, the autocorrelation of \(m_{\eta_c}\) is significantly suppressed with increasing \(\tau\), showing a marked improvement from \(\tau=0.5\) to \(\tau=1.0\). In contrast, a larger \(\tau\) provides only a minor reduction in the autocorrelation of \(\mathrm{ReTr}(P)\). Both \(Q\) and \(m_{\pi}\) demonstrate much weaker autocorrelations, remaining negligible even at the finest lattice spacing of \(a = 0.047\) fm.

\begin{table}[h]
\centering
\caption{$a_{w_0}$ on different ensembles and the fitted values through the functional form in Eq.~(\ref{eq:a_global_fit}). The values in the column labeled by $a_{w_0}$ only show the statistical errors. Other fitted results include the uncertainty from $w_0=0.17250(70)$ fm. }
\begin{tabular}{l|l|l||c|c}
 & $a_{w_0}$~(fm) & $a^{\rm fit}_{w_0}$~(fm) & \multicolumn{2}{c}{fit parameters}  \\
\hline
c24P31s & $0.11121(14)$ & $0.11098(46)$ & $a(7.29)$ & 0.10840(44) \\
c24P31 & $0.11089(14)$ & $0.11089(47)$ & $a(7.54)$ & 0.08668(37) \\
c24P22 & $0.10938(15)$ & $0.10944(45)$ & $a(7.75)$ & 0.07103(30) \\
c32P31 & $0.11094(08)$ & $0.11100(46)$ & $a(8.20)$ & 0.04730(30) \\
c32P22 & $0.10943(09)$ & $0.10940(45)$ & $c_{l}$   & 0.311(17)   \\
c48P13 & $0.10843(05)$ & $0.10843(44)$ & $c_{s}$   & -0.022(18)  \\
e32P31 & $0.08880(11)$ & $0.08879(38)$ & $c_{u_0}$ & -5(12)      \\
g32P31 & $0.07276(14)$ & $0.07283(30)$ & 
\\
g48P31 & $0.07278(07)$ & $0.07276(30)$ &  &\\
h48P31 & $0.04845(22)$ & $0.04845(30)$ \\
\end{tabular}
\label{tab:lat_a_fit}
\end{table}

We determine the lattice spacing \(a_{w_0}\) for each ensemble using the gradient flow scale with \(w_0 = 0.1725(7)\) fm, based on the FLAG average~\cite{FlavourLatticeAveragingGroupFLAG:2024oxs,Dowdall:2013rya,ExtendedTwistedMass:2021qui,Miller:2020evg,MILC:2015tqx,Borsanyi:2020mff}. The results are summarized in Table~\ref{tab:lat_a_fit}. To perform a global fit, we adopt the empirical formula from Ref.~\cite{CLQCD:2023sdb}, originally used for CLQCD clover fermion ensembles. The fit function is:
\begin{align}
&a_{w_0}(\hat{\beta},\tilde{m}_{\pi}, \tilde{m}_{\eta_s},\delta u_0)=a(\hat{\beta})\big[1+\nonumber\\
&\quad\quad\quad + c_l(\frac{\tilde{m}_{\pi}^2}{a(\hat{\beta})^2}-m_{\pi,{\rm phys}}^2)+c_s(\frac{\tilde{m}_{\eta_s}^2}{a(\hat{\beta})^2}-m_{\eta_s,{\rm phys}}^2)\nonumber\\
&\quad\quad\quad + c_{u_0}(u_0-u_0^I)\big]\;,\label{eq:a_global_fit}
\end{align}
Here, \(m_{\pi,{\rm phys}}=134.98\) MeV~\cite{ParticleDataGroup:2020ssz} is the physical pion mass with QED corrections highly suppressed, \(m_{\eta_s,{\rm phys}}=689.63(18)\) MeV~\cite{Borsanyi:2020mff} corresponds to the physical strange quark mass, and \(u^I_0\) denotes the input value of $u_0$. The fit yields a \(\chi^2\)/d.o.f. of 1.3, and the resulting parameters are also provided in Table~\ref{tab:lat_a_fit}. 

To isolate the effect of the charm sea quark and gauge actions, we also generated three 2+1-flavor ensembles with a light-to-strange mass ratio of $\tilde{m}^{\rm b}_{s}/\tilde{m}^{\rm b}_{l}=5$: x24P31 using the S$^{(0)}$ gauge action, y24P31 using the S$^{\rm tad}$ action gauge, and z24P31 using the Iwasaki gauge actions. The parameters $\tilde{m}^{\rm b}_{s}$ and $\hat{\beta}$ were tuned to match both the $m_{\eta_s}$ and the lattice spacing of the 2+1+1-flavor ensemble c24P30, enabling a direct comparison of quantities with and without charm sea quarks. 

\begin{table}[!h]
\centering
\caption{Input parameters $\hat{\beta}$, $u_0$, and $\tilde{m}^{\rm b}_{l,s}$ of the 2+1 flavor gauge ensembles with size $\tilde{L}\times \tilde{T}=24^3\times 48$ HISQ fermion and different gauge actions.}
\begin{tabular}{c|clc|ll}
\hline
Ensemble & $C_R$ & $\hat{\beta}$ & $u_0$ &$\tilde{m}^{\rm b}_{l}$ & $\tilde{m}^{\rm b}_s$ \\
\hline 
x24P31 &-1/20 &6.743&0.872958(4)&0.01027&0.05135\\
y24P31 &-1/(20$u_0^2$) &7.2133&0.877265(3)&0.009796& 0.04898\\
z24P31 &-0.0907346 &8.322&0.886683(3)&0.010582&0.05291\\
\hline
\end{tabular}
\label{tab:hisq_inputs2}
\end{table}

The input parameters of the 2+1 flavor ensembles are collected in Table~\ref{tab:hisq_inputs2}. Based on fitted value of the coefficient \(c_l\), the lattice spacing \(a_{w_0}\) exhibits a -2\% shift when moving from \(m_{\pi} \sim 300\) MeV to the physical point. We therefore apply a 0.98 rescaling factor to \(a_{w_0}\) for the x(y/z)24P31 ensembles. Given that $\Delta_{\rm mix}$ has mass dimension 2, we assign a 4\% systematic uncertainty to its value for the x(y/z)24P31 ensembles to account for this rescaling.

Furthermore, we observe that $u_0$ is comparable for all three ensembles, despite considerable variation in $\hat{\beta}$. This feature is expected, as the effective gauge coupling defined through tadpole improvement,
\begin{align}
\alpha^{\rm tad}_s\equiv \frac{6(1-40/3C_Ru_0^2)}{4\pi\hat{\beta}u_0^4},
\end{align}
is found to be similar at the 2–3\% level for all the three $C_R$ values used in this work.

Using the rescaled lattice spacing, the $\eta_s$ meson mass $m_{\eta_s}$ for both the x24P31 and z24P31 ensembles is found to be 2-3\% larger than the value corresponding to the physical strange quark mass. The resulting rescaled strange quark masses are 0.0483(3) and 0.0508(3), respectively. These values show a clear increase with the gauge coupling $\hat{\beta}$ across all three ensembles studied.

\begin{table}[!h]
\centering
\caption{Summary table on the lattice spacing, $\tilde{L}\times \tilde{T}$, $m_{\pi}L$  and $m_{\pi,\eta_s,\eta_c}$ (In unit of MeV) of the CLQCD ensembles with HISQ fermion. }
\begin{tabular}{l|lcc|ccc}
\hline
Ensemble & $a$ (fm) & $\tilde{L}^3 \times \tilde{T}$ &$m_{\pi}L$ & $m_{\pi}$ &$m_{\eta_s}$ & $m_{\eta_c}$ \\
\hline 
c24P31s &\multirow{6}[2]{*}{0.1084(4)}&$24^3\times 48$&4.13&313(2)&745(3)&2.973(12)\\
c24P31 &&$24^3\times 48$&4.07&309(2)&687(3)& 2.972(12)\\
c32P31 &&$32^3\times 48$&5.44&310(1)&686(3)&2.972(12)\\
c24P22 &&$24^3\times 48$&2.94&223(2)&685(3)&2.970(12) \\
c32P22 &&$32^3\times 48$&3.87&220(1)&684(3) &2.970(12)\\
c48P13 &&$48^3\times 48$&3.53&134(1)&683(3)& 2.970(12)\\
\hline
e32P31 &0.0867(4)&$32^3\times 64$&4.41&313(2)&694(3)&3.015(13)\\
\hline
g32P32 &\multirow{2}[2]{*}{0.0710(3)}&$32^3\times 64$&3.65&317(3)&692(3)&2.981(13)\\
g48P31 &&$48^3\times 64$&5.38&311(2)&691(3)&2.983(13)\\
\hline
h48P31 &0.0473(3)&$48^3\times 96$&3.60&313(3)&692(5)&2.947(19)\\
\hline
\hline
x24P31&0.1114(6)&$24^3\times 48$&4.35&321(2)&711(4)&--\\
y24P31&0.1116(6)&$24^3\times 48$&4.22&311(2)&688(4)&--\\
z24P31&0.1128(7)&$24^3\times 48$&4.35&317(2)&704(4)&--\\
\hline
\end{tabular}
\label{tab:hisq_ensembles}
\end{table}

Table~\ref{tab:hisq_ensembles} lists the pseudoscalar meson masses $m_{\pi}$, $m_{\eta_s}$, and $m_{\eta_c}$, determined using the jointly fitted lattice spacing. The $m_{\eta_s}$ values on all the 2+1+1 flavor ensembles are consistent with the our target value of 689.63(18) MeV~\cite{Borsanyi:2020mff} within $2\sigma$, except for the c24P31s ensemble used to test strange mass mistuning. The $m_{\eta_c}$ results also reproduce the physical value to within approximately 1\%.

Using Table~\ref{tab:hisq_ensembles}, we perform a second-order polynomial interpolation (or extrapolation) to determine the values of $\hat{\beta}$, $u_0$, and $\tilde{m}_s^{\rm b}$ at the lattice spacings of the $N_f=2+1$ CLQCD ensembles. These interpolated values are collected in the appendix.

\begin{table}[hbt!]
\centering
\caption{Comparison of the lattice spacing \(a\), tadpole factors \(u_0\) and \(v_0^{\rm(HYP)}\), and the bare light quark mass \(m_l^{\rm b}\) (for \(m_{\pi} \simeq 300\) MeV) using the HYP-smeared clover action. Symbols: HI (HISQ), SC (stout-smeared clover), S\(^{(0)}\) (tree-level Symanzik), S\(^{\rm tad}\) (tadpole-improved Symanzik), S\(^{(1)}\) (one-loop-improved Symanzik), I (Iwasaki).}
\begin{tabular}{lc|cc|ccc}
\hline
Action & $N_f$ & $\hat{\beta}$ & $a$ (fm) &  $u_0$ &$v_0^{(\rm HYP)}$& $\tilde{m}_l^{\rm b(HYP)}$\\
\hline 
\multirow{4}[2]{*}{HI+S$^{\rm tad}$} & \multirow{4}[2]{*}{2+1+1} 
&  7.29 & 0.1084(04) & 0.87944 & 0.9862 & -0.0529
\\
&& 7.54 & 0.0867(03) & 0.88636 & 0.9878 & -0.0438 \\
&& 7.75 & 0.0710(03) & 0.89143 & 0.9889 & -0.0397 \\
&& 8.20 & 0.0473(03) & 0.90060 & 0.9905 & -0.0329 \\
\hline
HI+S$^{(0)}$ & 2+1 
& 6.74 & 0.1114(06) & 0.87296 &0.9852 &-0.0578  \\
\hline
HI+S$^{\rm tad}$ & 2+1 
&  7.21 & 0.1116(06) & 0.87727 &0.9858 &-0.0558 \\
\hline
HI+I & 2+1 
& 8.32 &  0.1128(07) & 0.88668 &0.9870 &-0.0485  \\
\hline
\multirow{3}[2]{*}{SC+S$^{\rm tad}$} & \multirow{3}[2]{*}{2+1} 
&  6.20 & 0.1052(06) & 0.85545 & 0.9830 & -0.0328 \\
&& 6.41 & 0.0775(05) & 0.86346 & 0.9851 & -0.0208 \\
&& 6.72 & 0.0520(03) & 0.87338 & 0.9871 & -0.0135 \\
\hline
\multirow{4}[2]{*}{HI+S$^{(1)}$} & \multirow{4}[2]{*}{2+1+1} 
&  6.00 & 0.1222(03) & 0.86373 & 0.9836 & -0.0708 \\
&& 6.30 & 0.0879(02) & 0.87417 & 0.9863 & -0.0514 \\
&& 6.72 & 0.0566(01) & 0.88578 & 0.9887 & -0.0398 \\
&& 7.00 & 0.0426(01) & 0.89218 & 0.9897 & -0.0365 \\
\hline
\end{tabular}
\label{tab:action_comapre}
\end{table}

Table~\ref{tab:action_comapre} summarizes the parameters $\hat{\beta}$, $a$, and tadpole factors \(u_0\) and \(v_0^{\rm(HYP)}\) for the following ensembles:

\begin{itemize}
    \item \textbf{HI+S\(^{\rm tad}\)}: The 2+1+1 and 2+1 flavor HISQ fermion ensembles with the tadpole-improved Symanzik gauge action generated in this work;
    \item \textbf{HI+S\(^{(0)}\)} or \textbf{HI+I}: The 2+1 flavor HISQ fermion ensembles with the tree-level Symanzik (or Iwasaki) gauge action generated in this work;
    \item \textbf{SC+S\(^{\rm tad}\)}: The 2+1 flavor stout-smeared clover (SC) fermion ensembles with the same S\(^{\rm tad}\) gauge action from the CLQCD collaboration~\cite{CLQCD:2023sdb};
    \item \textbf{HI+S\(^{(1)}\)}: The 2+1+1 flavor HISQ fermion ensembles with a one-loop-improved Symanzik (S\(^{(1)}\)) gauge action from the MILC collaboration~\cite{MILC:2010pul, MILC:2012znn} (data taken from Ref.~\cite{Zhao:2022ooq}).
\end{itemize}

\begin{figure}[hbt!]
\centering
\includegraphics[width=0.48\textwidth]{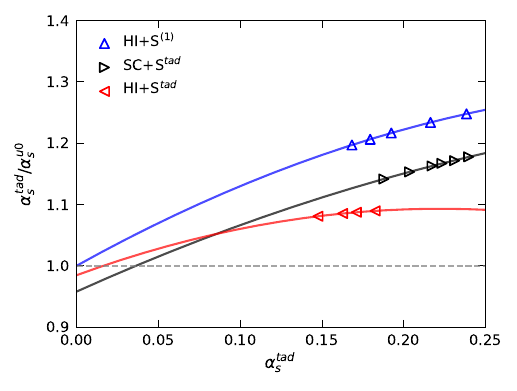}
\caption{The ratio $\alpha^{\rm tad}_s/\alpha^{u_0}_s$ of two definitions of effective $\alpha_s$ using different fermion and gauge action combinations. they approach to 1 within a few percent deviations in the weak coupling limit $\alpha_s\rightarrow 0$.}
\label{fig:autocorr}
\end{figure}

A comparison reveals that at a similar lattice spacing of $a = 0.1116$ fm, the value of $\hat{\beta}$ is 0.6\% larger for the 2+1+1 flavor ensembles than for the 2+1 flavor case. This indicates that the tadpole improved bare strong coupling $\alpha^{\rm tad}_s$ decreases by approximately 1.1\% when charm sea quark effects are included. Using the same formulation, we find that $\alpha_s^{\rm tad}$ obtained with the 2+1 flavor HISQ action is 1.2\% and 2.8\% smaller when paired with the tree-level Symanzik or Iwasaki gauge action, respectively, compared to the result with the tadpole-improved Symanzik gauge action.
In contrast, changing either the fermion action from SC to HI or the gauge action from S\(^{\rm tad}\) to S\(^{(1)}\) leads to a much more significant shift in $\hat{\beta}$, corresponding to an $\sim 18$\% change in $\alpha^{\rm tad}_s$.

For the Symanzik gauge action, the strong coupling $\alpha_s$ can be estimated at leading-order perturbation theory through the tadpole factor $u_0$~\cite{Alford:1995hw}, defined as
\begin{align}
\alpha^{u_0}_s \equiv -\frac{4}{3.06839} \log u_0.
\end{align}
As illustrated in Fig.~\ref{fig:autocorr}, the ratio $\alpha^{\rm tad}_s / \alpha^{u_0}_s$ depends on the specific combination of fermion and gauge actions. However, this ratio consistently approaches unity with deviations of only a few percent in the weak coupling limit $\alpha_s \to 0$, using extrapolation with a polynomial form $\sum_{i=0}^2c_i\alpha_s^i$.

As shown in Table~\ref{tab:action_comapre}, we observe that the HYP-smeared tadpole factor \(v_0^{(\rm HYP)}\) deviates from unity by less than 2\% in all ensembles, with a scaling that is faster than \(\alpha_s(a)\). This implies that \({\cal O}(a^n)\) discretization effects can be as important as \(\alpha_s(a)\) corrections. The same logic applies to the dimensionless bare quark mass \(\tilde{m}_l^{\rm b}\equiv m_l^{\rm b}a\) which corresponds to $m_{\pi}\simeq 300$ MeV, and its scaling provides clear evidence for both the \(\alpha_s^n/a\) divergence and \({\cal O}(a^n)\) discretization effects in the clover fermion's bare mass parameter.

\section{Result of $\Delta_{\rm mix}$}\label{sec:setup}

We calculate the leading mixed-action effects $\Delta_{\text{mix}}$ on 2+1+1 flavor gauge ensembles c24P31, e32P31, g32P31 and h48P31  at four lattice spacings with $m_{\pi}\sim 310$ MeV, and also 2+1 flavor gauge ensembles x24P31, y24P31 and z24P31 with different gauge actions. For the valence fermion actions, we uses:

\begin{itemize}
    \item \textbf{SC}: Clover fermion with 1-step STOUT smearing and tree level tadpole improved clover coefficient $c_{\text{sw}}$;
    \item \textbf{HC}: Clover fermion with 1-step HYP smearing and tree level tadpole improved clover coefficient $c_{\text{sw}}$;
    \item \textbf{OV}: Overlap fermion with 1-step HYP smearing and $\rho=1.5$.
\end{itemize}

Even-grid source (including $\tilde{L}^3/8$ points at all the spatial positions where the x/y/z indices are even) are used generate propagator for all the valence fermion actions to improve the statistics. In practice, on each ensemble we compute quark propagators for the sea (HISQ) action and for the valence action at several valence quark masses $m_v$. We construct pseudoscalar two-point functions for the sea–sea (ss), valence–valence (vv), and valence–sea (vs) channels and extract $m_{\pi,ss}$, $m_{\pi,vv}(m_v)$, and $m_{\pi,vs}(m_v)$ from standard fits. We then interpolate in $m_v$ to the matching point where $m_{\pi,vv}=m_{\pi,ss}$, and evaluate $\Delta_{\mathrm{mix}}$ using Eq. (1) at this point. Since our sea action employs HISQ fermions, calculating $\Delta_{\text{mix}}$ necessitates an expansion of the Dirac indices for the corresponding HISQ quark propagator $S^{\text{HISQ}}$. For this purpose, we extend $S^{\text{HISQ}}$ into unitary matrices in the Dirac space, and multiply products of gamma matrices at each sink point $(x_{\text{snk}},y_{\text{snk}},z_{\text{snk}},t_{\text{snk}})$ of the even-grid source propagator, namely $S^{\text{Dirac}} = \gamma_t^{t_{\text{snk}}}\gamma_z^{z_{\text{snk}}}\gamma_y^{y_{\text{snk}}}\gamma_x^{x_{\text{snk}}} S^{\text{HISQ}}$.


All our $2+1+1$ flavor results of $\Delta_{\text{mix}}$ are collected in Table~\ref{tab:hisq_mix}, together with a generalized mixed action effect between valence fermion actions~\cite{Zhao:2022ooq},
\begin{equation}
\bar{\Delta}_{\text{mix}}^{\text{B+C/A}} \equiv m_{\pi,\text{BC}}^{2} - \frac{m_{\pi,\text{BB}}^{2} + m_{\pi,\text{CC}}^{2}}{2}|_{m_{\pi,\text{AA}}=m_{\pi,\text{BB}}=m_{\pi,\text{CC}}},
\end{equation}
which quantifies the effect of using two different valence actions B and C on the same sea action A. For computational efficiency, we calculate $\Delta_{\text{mix}}$ using the overlap valence fermion action only at two representative lattice spacings, $a \sim 0.11$ and 0.07 fm. 

\begin{figure}[!h]
\centering
\includegraphics[width=0.50\textwidth]{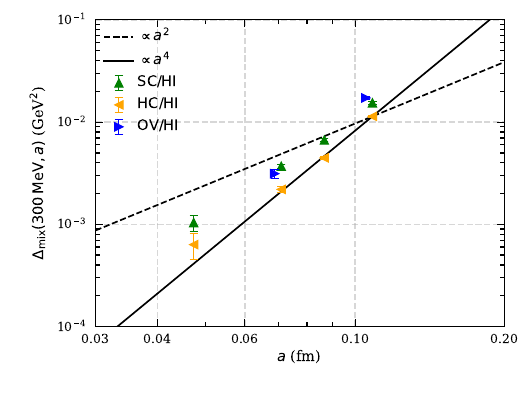}
\caption{Mixed-action effects of the Stout link smeared Clover fermion (SC/HI, green triangles) and  HYP smeared Clover fermion (HC/HI, yellow left triangles), overlap fermion (OV, blue right triangles, at a=0.108 and 0.073 fm only) on the 2+1+1 flavor HISQ ensembles at different lattice spacings.}
\label{fig:DELTA}
\end{figure}

\begin{table*}[ht!]                   
\caption{$\Delta_{\rm mix}$ and $\bar{\Delta}_{\rm mix}$ on different ensembles with different valence quark actions. The symbol OV, SC and HC correspond to the overlap, stout-smeared clover, and HYP smeared clover fermion actions.}\label{tab:mix_act}
\begin{tabular}{c c c c| c c c c c c}                                      
\multicolumn{4}{c}{} & \multicolumn{3}{c}{$\Delta_{\rm mix}$(\text{GeV}$^2$)} & \multicolumn{3}{c}{$\bar{\Delta}_{\rm mix}$(\text{GeV}$^2$)} \\
\text{Action} &\text{Symbol} &\text{a(fm)} &$m_{\pi,{\rm ss}}$ (MeV) &\text{OV} &\text{SC} &\text{HC} &\text{OV+SC} &\text{OV+HC} &\text{SC+HC}   \\
\hline   
\multirow{4}[2]{*}{2+1+1} & c24P30 &0.108&309 &0.01724(76) & 0.01545(45) & 0.01138(35)& 0.01344(54) & 0.00700(40) & 0.00292(12)\\  
 & e32P31&0.087&313 &- & 0.00668(20) & 0.00447(15)& - & - & 0.00156(06) \\  
 & g32P31 &0.071&317&0.00313(31)&0.00370(17)& 0.00220(13)& 0.00321(28)& 0.00126(23)& 0.00100(06)\\ 
 & h48P31 &0.047&313&- & 0.00104(18)& 0.00063(18) &-& - & 0.00027(03)\\  
\hline
\multicolumn{4}{c|}{$c_2$ (GeV$^4$)} & -0.001(05) & 0.010(02) & 0.001(02) & 0.010(04) & -0.001(03) & 0.005(01) \\
\multicolumn{4}{c|}{$c_4$ (GeV$^6$)} & 0.196(21) & 0.135(10) & 0.120(08) & 0.118(17) & 0.080(13) & 0.181(03) \\
\hline
\multirow{3}[3]{*}{2+1} & x24P31&0.111&321&0.02366(80)&0.01965(50)& 0.01449(41)&0.01787(65)&0.00952(43) & 0.00383(15)\\
& y24P31 &0.112&311&0.01997(82) &0.01725(49)& 0.01259(36)& 0.01563(93)& 0.00902(76)& 0.00325(15)\\
& z24P31 &0.113&317& 0.01378(59) &0.01272(35)& 0.00937(26)&0.01113(41)&0.00554(30)& 0.00245(08)\\
\hline
\end{tabular}  
\label{tab:hisq_mix}
\end{table*}

The lattice spacing dependence of $\Delta_{\text{mix}}$ on the 2+1+1 flavor HISQ ensembles is shown in Fig.~\ref{fig:DELTA}. The solid (dashed) line passes through the HC/HI $\Delta_{\mathrm{mix}}$ point at the coarsest spacing and indicates $a^4$ ($a^2$) scaling. The SC and HC results exhibit a clear $\mathcal{O}(a^4)$ scaling across the four lattice spacings, despite a large relative uncertainty at the finest lattice spacing ($a = 0.048$ fm). The overlap results, though limited to two data points, are also consistent with this $\mathcal{O}(a^4)$ scaling behavior. We further parameterize $\Delta_{\text{mix}}$ with the following form,
\begin{align}\label{eq:delta_fit}
    \Delta_{\text{mix}}(a)=c_2a^2+c_4a^4,
\end{align}
and list the fitted (or solved in the overlap case) coefficients $c_{2,4}$ in Table~\ref{tab:hisq_mix}. It is evident that the values of $c_2$ are significantly non-zero in cases involving the stout-link smeared clover action (SC), although they remain much smaller than the corresponding $c_4$ values. Since \(a^4\) scaling is observed for both the exactly chiral overlap fermion and the HYP-smeared clover fermion (with its suppressed chiral symmetry breaking), the scaling can be strongly related to chiral symmetry.

\begin{figure*}[hbt!]
\centering
\includegraphics[width=0.80\textwidth]{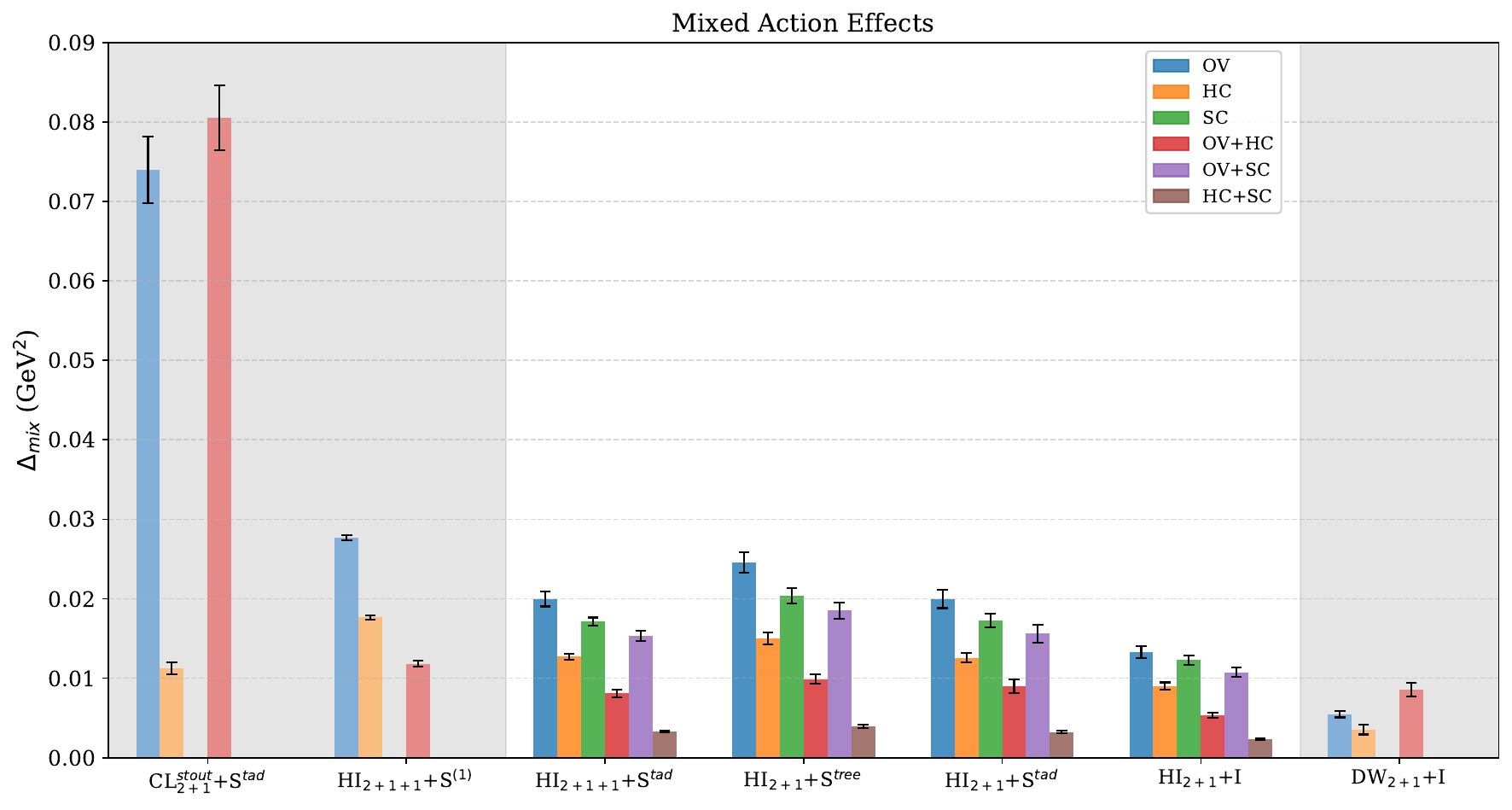}
\caption{The mixed action effects $\Delta_{\rm mix}$ at $m_{\pi}\sim 0.30$ GeV for various valence fermion actions (OV, SC and HC) and $\bar{\Delta}_{\rm mix, uni}$ for different valence fermion actions (OV+HC, OV+SC and HC+HC) on the 2+1 flavor HI+I, HI+S$^{\rm tad}$, HI+S$^{(0)}$ ensembles, 2+1+1 flavor HI+S$^{\rm tad}$ ensembles, and also includes the values from Ref.~\cite{Zhao:2022ooq} (shown as histograms in the shaded region) for comparison. All the quantities are in unit of GeV$^2$, and are interpolated to $a=0.112$ fm and $m_{\pi}\sim 300$ MeV to make a fair comparison.}
\label{fig:mix_compare}
\end{figure*}

Given that the lattice spacings of the x/z24P31 ensembles ($a$=0.111/0.113 fm) differ from that of the y24P31 ensemble ($a$=0.112 fm) by only 1\%, we interpolated all the $2+1+1$ flavor results to a common reference point of $a=0.112$ fm using the parameterization in Eq.~(\ref{eq:delta_fit}). This enables a direct comparison of $\Delta_{\text{mix}}$ across different fermion-gauge action combinations and charm quark content. The interpolated results are presented in Fig.~\ref{fig:mix_compare}.

A comparison between the 2+1 and 2+1+1 flavor HISQ ensembles with the same S$^{\rm tad}$ gauge action reveals that $\Delta_{\text{mix}}$ is consistent within statistical uncertainties, regardless of the inclusion of charm sea quarks. In contrast, both of these $\Delta_{\text{mix}}$ values are lower than the values~\cite{Zhao:2022ooq} on the 2+1+1 flavor HISQ ensemble with the S$^{(1)}$ gauge action. 

For the 2+1 flavor cases, $\Delta_{\text{mix}}$ exhibits an obvious decrease with increasing absolute value of $|C_P|$, rendering its value with the Iwasaki action being 0.61(4) of that with the tree level Symanzik gauge actions. Nevertheless, $\Delta_{\text{mix}}$ with DW fermions is still smaller by a factor of 2.5(3) compared with HISQ fermions under the same Iwasaki gauge action. When using the same S$^{\rm tad}$ gauge action, $\Delta_{\text{mix}}$ with HISQ fermions is significantly smaller than with stout clover (SC) fermions, confirming the importance of using a chirally symmetric fermion action for generating gauge configurations on suppression $\Delta_{\text{mix}}$.

The preceding calculations enable us to isolate the specific origins of variations in $\Delta_{\text{mix}}$ by comparing our results with existing literature:

1.  OV@TM$_2$+S$^{\text{tree}}$~\cite{Cichy:2012vg}: Using ensembles with 2-flavor twisted mass sea quarks and a tree-level Symanzik gauge action at $m_{\pi}\in [376,484]$ MeV and $a\in[0.05,0.08]$ fm, the reported $\Delta_{\text{mix}}$ for a 1-step HYP-smeared overlap valence fermion ($\rho=1$) is 0.34(4) GeV$^2$ when extrapolated to $a=0.112$ fm using an $a^2+a^4$ ansatz Eq.~(\ref{eq:delta_fit}). The dominant scaling was $a^2$, with an $a^4$ term coefficient of $c_4=0.4(6)$ GeV$^6$. This value is substantially larger than the 0.071(4) GeV$^2$ found for OV@CL$_{2+1}$+S$^{\text{tad}}$~\cite{Zhao:2022ooq}. This significant difference is likely attributable to the different sea fermion actions, as the effects from the number of sea flavors and the gauge action are probably insufficient to explain it.

2.  OV@CL$_{2+1}$+S$^{\text{tree}}$~\cite{Durr:2007ef}: Using a 2+1 flavor ensemble with 6-step stout-smeared clover sea quarks and a tree-level Symanzik gauge action at $m_{\pi}\in [190,300]$ MeV and $a=0.088$ fm, $\Delta_{\text{mix}}$ for a 3-step HYP-smeared overlap valence fermion ($\rho=1$) was 0.12(5) GeV$^2$. Scaling to $a=0.112$ fm assuming $a^2$ dependence gives 0.19(8) GeV$^2$. This is consistent with the 0.071(4) GeV$^2$ from OV@CL$_{2+1}$+S$^{\text{tad}}$~\cite{Zhao:2022ooq} within the large uncertainty, especially considering the expected increase when switching from S$^{\text{tad}}$ to S$^{\text{tree}}$.

3.  OV@HI$_{2+1+1}$+S$^{(1)}$~\cite{Basak:2014kma}: Using a 2+1+1 flavor HISQ ensemble with a 1-loop Symanzik gauge action at $m_{\pi}= 310$ MeV and $a\simeq 0.12$ fm, $\Delta_{\text{mix}}$ for a 1-step HYP-smeared overlap valence fermion is 0.031(3) GeV$^2$ at $a=0.112$ fm, based on the $a^4$ scaling verified for this setup in Ref.~\cite{Zhao:2022ooq} (note: taste-breaking effects were unsubtracted~\cite{Basak:2017oup}). This value is in excellent agreement with the 0.027(3) GeV$^2$ from Ref.~\cite{Zhao:2022ooq}.

4.  DW@HI$_{2+1+1}$+S$^{(1)}$~\cite{Berkowitz:2017opd}: Using 2+1+1 flavor HISQ ensembles with a 1-loop Symanzik gauge action at $m_{\pi}\in [220,310]$ MeV and $a\in[0.09,0.15]$ fm, $\Delta_{\text{mix}}$ for a Mobius domain wall valence fermion is 0.024(1) GeV$^2$ at $a=0.112$ fm, based on an $a^2+a^4$ fit Eq.~(\ref{eq:delta_fit}) with coefficients $c_2=0.04(1)$ GeV$^{4}$ and $c_4=0.09(3)$ GeV$^{6}$. This is slightly smaller than the OV@HI$_{2+1+1}$+S$^{(1)}$ value, suggesting valence Mobius domain wall fermions may be a better choice on HISQ ensembles at coarser lattice spacings. However, the significantly non-zero $c_1$ term warrants further investigation.

5.  DW@AS$_{2+1}$+S$^{(1)}$~\cite{Orginos:2007tw,Aubin:2008wk,Walker-Loud:2008rui}: Using 2+1 flavor Asqtad-improved staggered (AS) sea quarks with a 1-loop Symanzik gauge action at $m_{\pi}\in [260,510]$ MeV and $a\in[0.09,0.13]$ fm, $\Delta_{\text{mix}}$ for a domain wall valence fermion is 0.065(5) GeV$^2$ at $a=0.112$ fm. The data were fit with an $a^2+a^4$ ansatz, but the uncertainty was too large to distinguish between the scaling terms. This value is roughly twice that of the most similar case (OV@HI$_{2+1+1}$+S$^{(1)}$), a difference we attribute to the AS sea action, given that domain wall and overlap fermions yield similar $\Delta_{\text{mix}}$ on HISQ ensembles.

6.  OV@DW$_{2+1}$+I~\cite{xQCD:2010pnl,Lujan:2012wg}: Using 2+1 flavor domain wall sea quarks with an Iwasaki gauge action at $m_{\pi}\in[300,400]$ MeV and $a\in[0.082,0.114]$ fm, $\Delta_{\text{mix}}$ for a 1-step HYP-smeared overlap valence fermion is 0.010(2) GeV$^2$ at $a=0.112$ fm. This is larger than the 0.0053(4) GeV$^2$ reported in Ref.~\cite{Zhao:2022ooq}, as the partially quenched chiral extrapolation used in Ref.~\cite{Lujan:2012wg} can about twice inflate the value.

7.  BC@AS$_{2+1}$+S$^{(1)}$~\cite{Basak:2017oup}: Using 2+1 flavor AS sea quarks with a 1-loop Symanzik gauge action at $m_{\pi}\sim 300$ MeV, $\Delta_{\text{mix}}$ for a Borici-Creutz valence fermion~\cite{Creutz:2007af,Borici:2007kz} was 0.017(6) GeV$^2$ at $a=0.15$ fm and consistent with zero (0.02(5) GeV$^2$) at $a=0.13$ fm. Considering the tendency of the AS sea action and S$^{(1)}$ gauge action to enlarge $\Delta_{\text{mix}}$, the Borici-Creutz fermion appears to be a promising candidate for mixed-action setups, though higher-precision calculations at finer lattice spacings are required for confirmation.


\section{Summary}\label{sec:summary}

In this work, we have performed a systematic lattice QCD calculation of the leading-order low-energy constant $\Delta_{\text{mix}}$ in mixed-action chiral perturbation theory. Our study utilized $2+1+1$ flavor HISQ sea fermions with a tadpole-improved Symanzik gauge action across four lattice spacings, and also $2+1$ flavor HISQ sea fermion with 3 kinds of gauge action at a fixed lattice spacing, enabling a controlled investigation into the sources of mixed-action artifacts.

Our central finding is the observation of a favorable $\mathcal{O}(a^4)$ scaling for $\Delta_{\text{mix}}$ when valence overlap or clover fermions are used on gauge ensembles generated with 2+1+1 flavor HISQ sea fermions. This result is particularly significant as it was achieved using the same tadpole-improved Symanzik gauge action previously associated with a sizable $\Delta_{\text{mix}}$ in CLQCD clover fermion ensembles~\cite{Zhao:2022ooq}. This comparison directly verifies that the use of a dynamical fermion action with chiral symmetry is the dominant factor in suppressing mixed-action artifacts, irrespective of the gauge action. Furthermore, by comparing different action combinations at a fixed lattice spacing, we have quantified additional, secondary influences on $\Delta_{\text{mix}}$. Our results show that the larger correction from rectangle loops in the gauge action can provide an additional suppression of $\Delta_{\text{mix}}$, while the effect of dynamical charm quarks is statistically insignificant.

A further comparison with values from the literature~\cite{Cichy:2012vg,Orginos:2007tw,Aubin:2008wk,Walker-Loud:2008rui} indicates that using either twisted mass or AS fermions as the sea action does not reduce mixed-action effects compared to the HISQ fermion. For the valence fermion, the domain wall fermion yields a $\Delta_{\text{mix}}$~\cite{Berkowitz:2017opd} comparable to that of either the clover or overlap fermion. If clover valence fermions are used, HYP smearing is preferable. The Borici-Creutz fermion appears to be a promising candidate for the valence fermion action~\cite{Basak:2017oup}, although higher-precision calculations at finer lattice spacings are required to confirm this.

In summary, our studies here that the HISQ sea fermion with the Iwasaki gauge action is the so far most cost-effective combination for suppressing $\Delta_{\text{mix}}$ with the flexibility in selecting valence fermion actions. If clover valence fermions are used, HYP smearing is preferable. The tadpole-improved Symanzik gauge action, while somewhat less effective at suppressing $\Delta_{\text{mix}}$, benefits from marginally weaker autocorrelations (see Appendix).
Further investigation on the Borici-Creutz fermion and different choice of the staggered fermion (likes the stout smearing with different steps and smearing sizes) would be helpful to further suppress $\Delta_{\text{mix}}$ and worth studies in the future.

Furthermore, a preliminary study~\cite{lin2025} of physical observable using clover valence fermions on our HISQ ensembles indicates that the discretization errors for quantities such as $f_{K}$ and $m_{\Omega}$ are suppressed compared to those from unitary clover fermion calculations~\cite{CLQCD:2023sdb,Hu:2024mas}. This suggests that the mixed-action setup can offer superior control over systematic uncertainty from continuum extrapolation, compared to the one with unitary valence-sea fermion action. Consequently, further investigations into a wider range of observables using this setup are warranted to identify optimal, cost-efficient strategies for lattice QCD calculations that maintain rigorous control over systematic errors.

\section*{Acknowledgement}
We thank the CLQCD collaborations for providing us their gauge configurations with dynamical fermions~\cite{CLQCD:2023sdb}, which are generated on HPC Cluster of ITP-CAS, IHEP-CAS and CSNS-CAS, the Southern Nuclear Science Computing Center(SNSC) and the Siyuan-1 cluster supported by the Center for High Performance Computing at Shanghai Jiao Tong University. 
The calculations were performed using the PyQUDA package~\cite{Jiang:2024lto} with QUDA~\cite{Clark:2009wm,Babich:2011np,Clark:2016rdz} through HIP programming model~\cite{Bi:2020wpt}. The numerical calculation were carried out on the ORISE Supercomputer, HPC Cluster of ITP-CAS and Advanced Computing East China Sub-center. This work is supported in part by National Key R\&D Program of China No.2024YFE0109800, NSFC grants No. 12435002, 12525504, 12293060, 12293062, and 12447101, the Strategic Priority Research Program of Chinese Academy of Sciences, Grant No.\ YSBR-101 and XDB34030303, and the science and education integration young faculty project of University of Chinese Academy of Sciences.

\bibliography{ref}

\clearpage
\onecolumngrid

\section*{Appendix}
\subsection{Autocorrelation Analysis}

To assess the statistical independence of gauge configurations generated by the Hybrid Monte Carlo algorithm, we perform an autocorrelation analysis using the binning method following the approach in Ref.~\cite{Du:2024}. The analysis focuses on four key observables: the average plaquette $\langle \text{Tr } U_p\rangle/3$, the topological charge $Q = \int d^4x q(x)$ with $q(x) = \frac{g^2}{32\pi^2}\epsilon_{\mu\nu\rho\sigma},\text{Tr},[F^{\mu\nu}(x)F^{\rho\sigma}(x)]$, the pion mass $m_\pi$, and the $\eta_c$ mass $m_{\eta_c}$. 

For a given observable $O$, the normalized variance ratio $\sigma_n/\sigma_1$ is computed as a function of bin size $n$. First, the overall mean is calculated as $\bar{O} = \frac{1}{N} \sum_{i=1}^{N} O_i$, where $O_i$ denotes an individual measurement and $N$ is the total number of configurations. For a bin size $n$, the data is partitioned into $M = \left\lfloor \frac{N}{n} \right\rfloor$ bins, and the mean for the $k$-th bin is given by $\bar{O}_{n,k} = \frac{1}{n} \sum_{i=(k-1)n+1}^{kn} O_i$.
The variance of these binned means is then computed as $\sigma_n^2(O) = \frac{1}{M} \sum_{k=1}^{M} \sigma_{n,k}^2(O)$ with $\sigma_{n,k}^2(O) = \frac{1}{M-1}(\bar{O}_{n,k} - \bar{O})^2$. To estimate the uncertainty of this variance, we calculate its standard error, $\mathrm{SE}(\sigma_n^2) = \sqrt{\frac{1}{M(M-1)}\sum_{k=1}^{M} \left(\sigma_{n,k}^2 - \sigma_n^2\right)^2}$.

Finally, the normalized ratio $\sigma_n/\sigma_1$ and its propagated error are evaluated. The error is given by $\delta\left(\frac{\sigma_n}{\sigma_1}\right) = \sqrt{\left(\frac{\delta\sigma_n}{\sigma_1}\right)^2 + \left(\frac{\sigma_n\delta\sigma_1}{\sigma_1^2}\right)^2}$, where $\delta\sigma_n = \mathrm{SE}(\sigma_n^2)/(2\sigma_n)$ and $\delta\sigma_1 = \mathrm{SE}(\sigma_1^2)/(2\sigma_1)$. This ratio quantifies the reduction in statistical error with increasing bin size, serving as our measure of autocorrelation.

\begin{figure}[htbp]
    \centering
    \begin{tabular}{ccc}
    \includegraphics[width=0.33\textwidth]{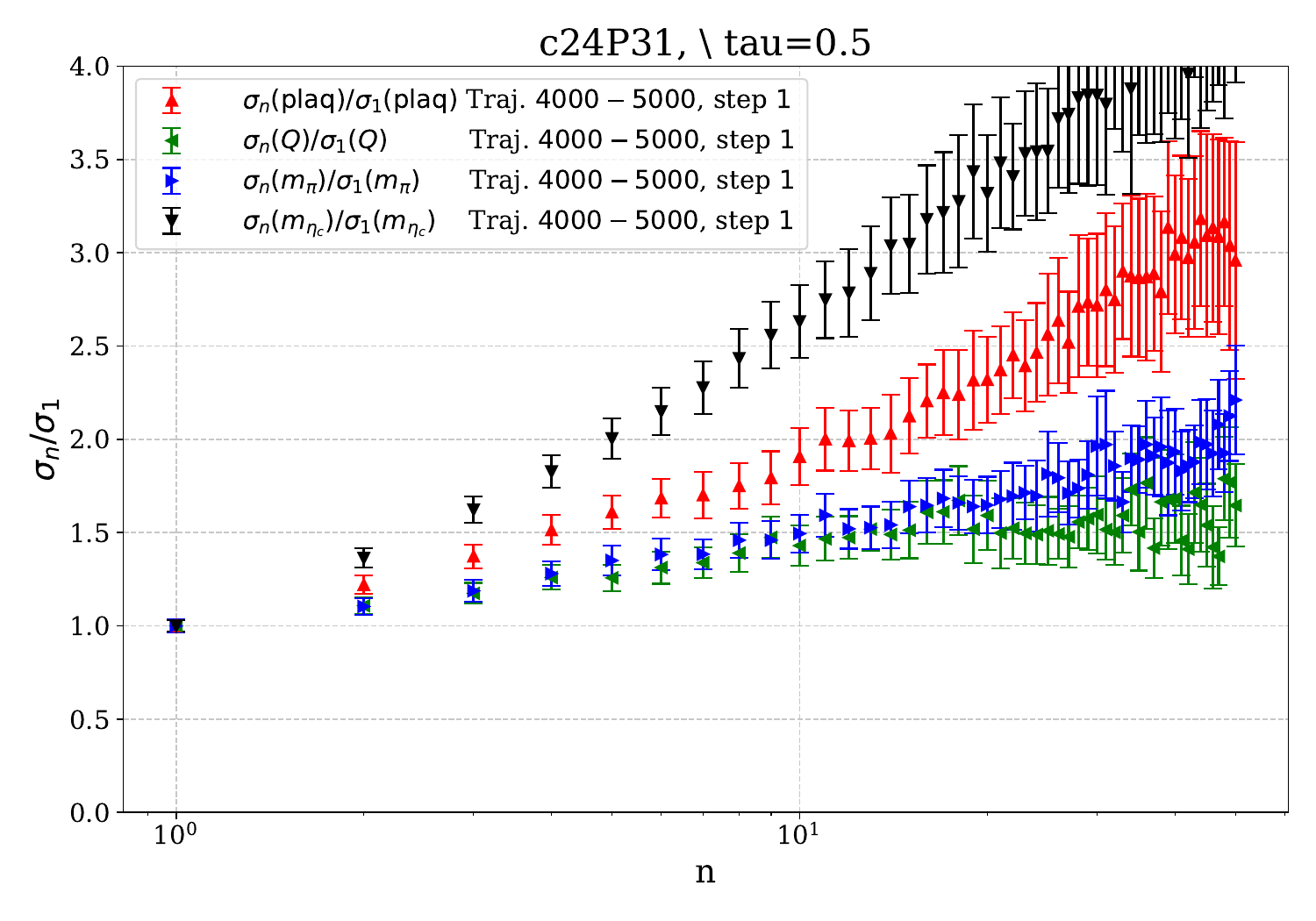} &
    \includegraphics[width=0.33\textwidth]{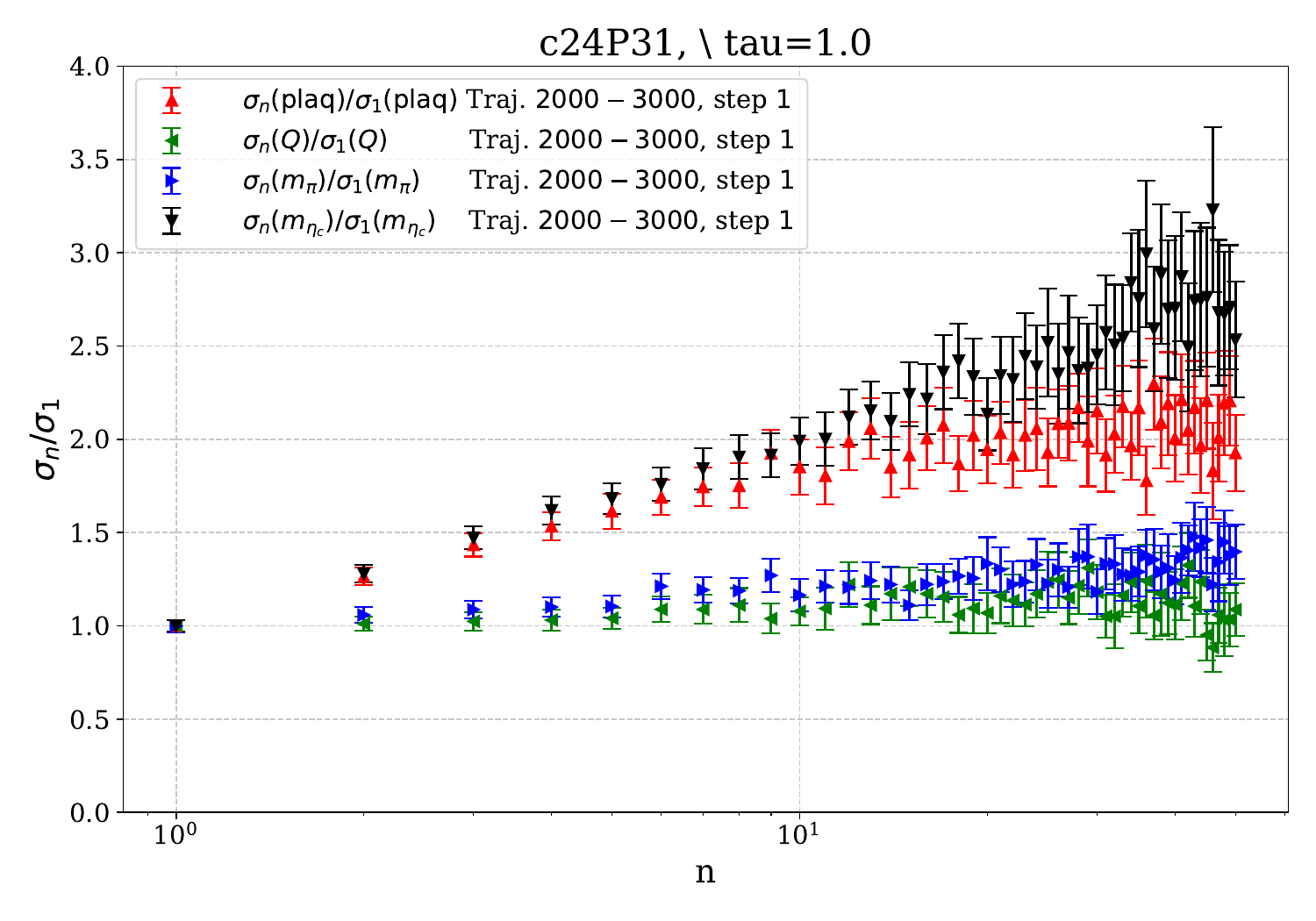} &
    \includegraphics[width=0.33\textwidth]{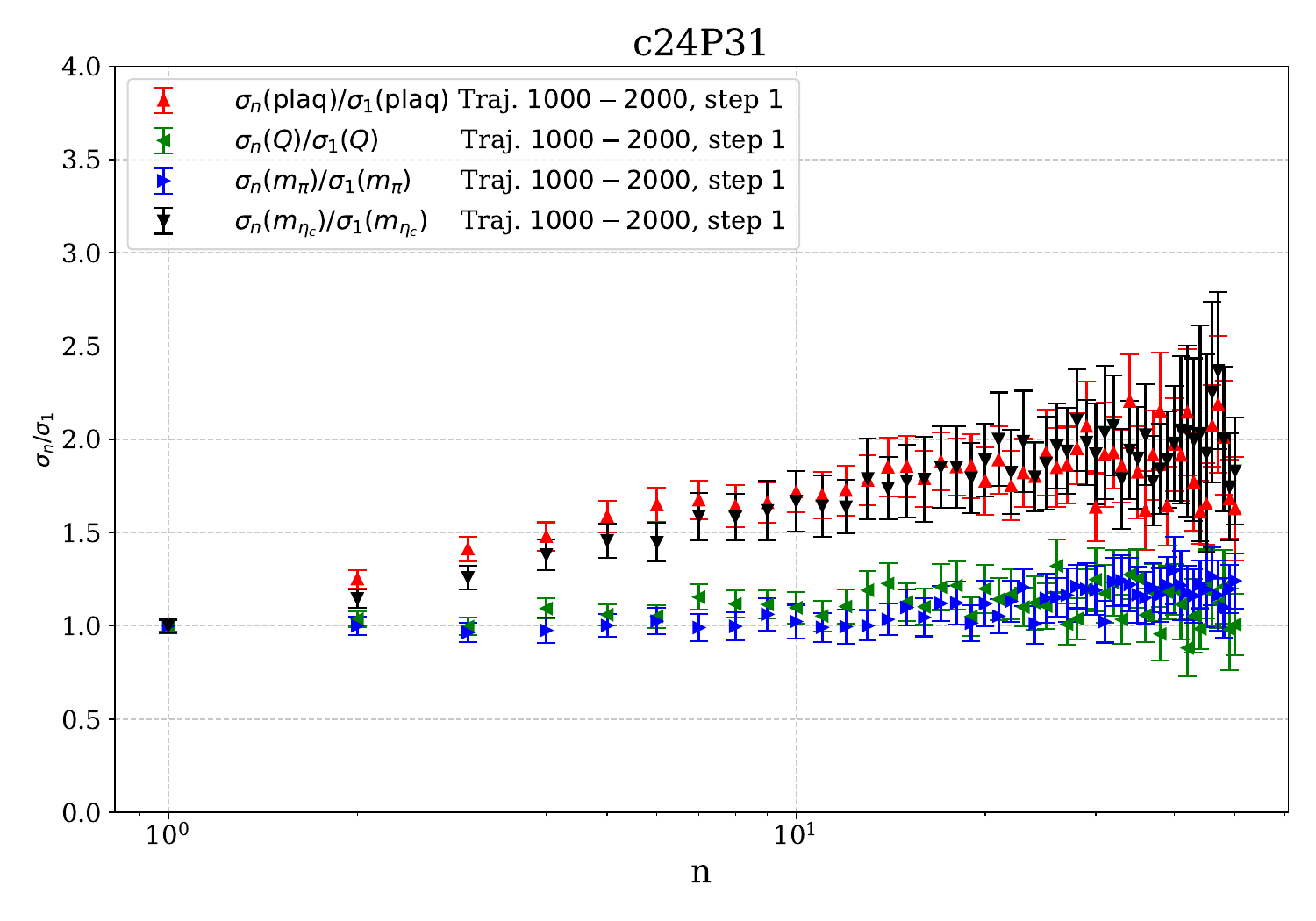} \\
    \includegraphics[width=0.33\textwidth]{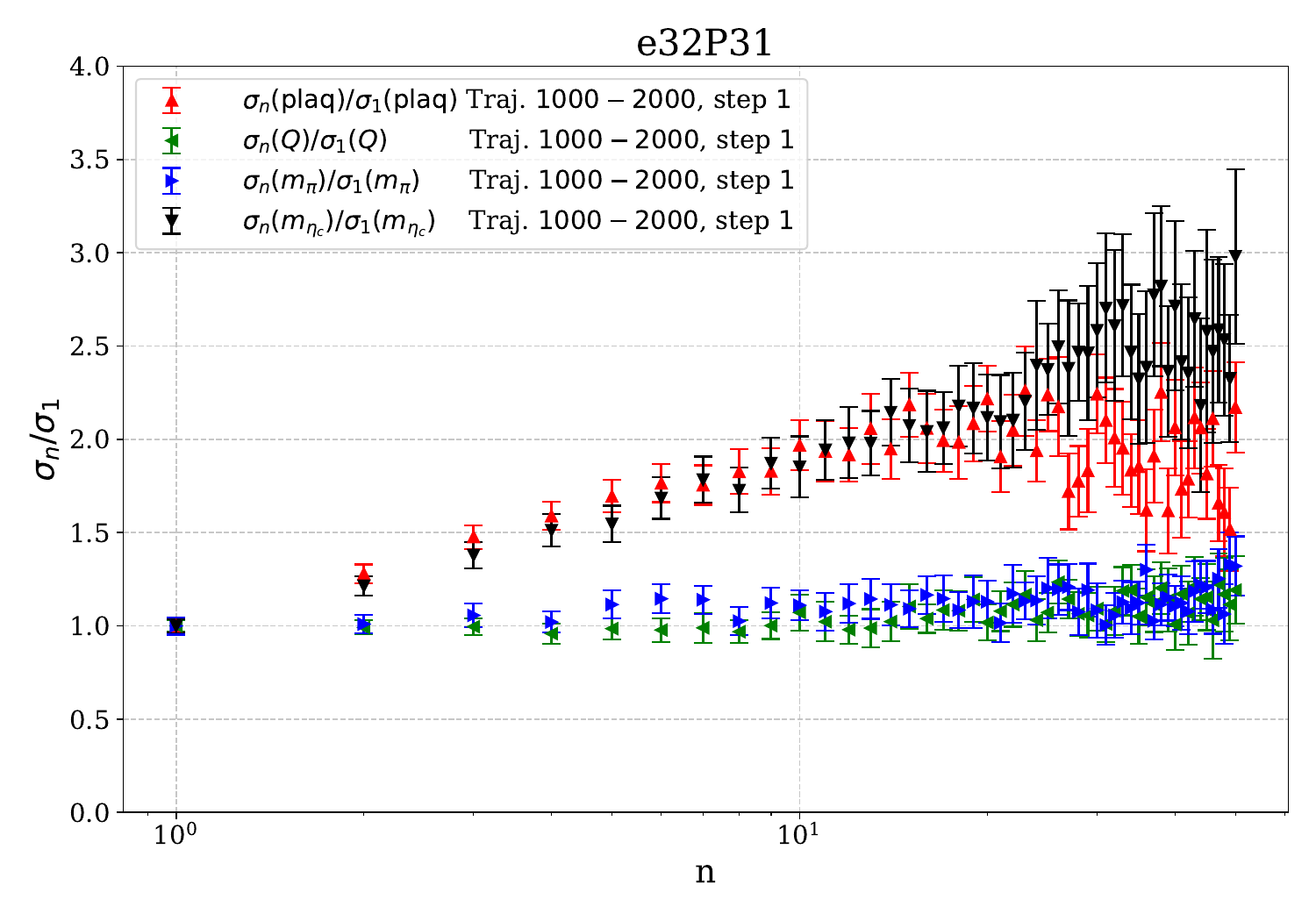}   &
    \includegraphics[width=0.33\textwidth]{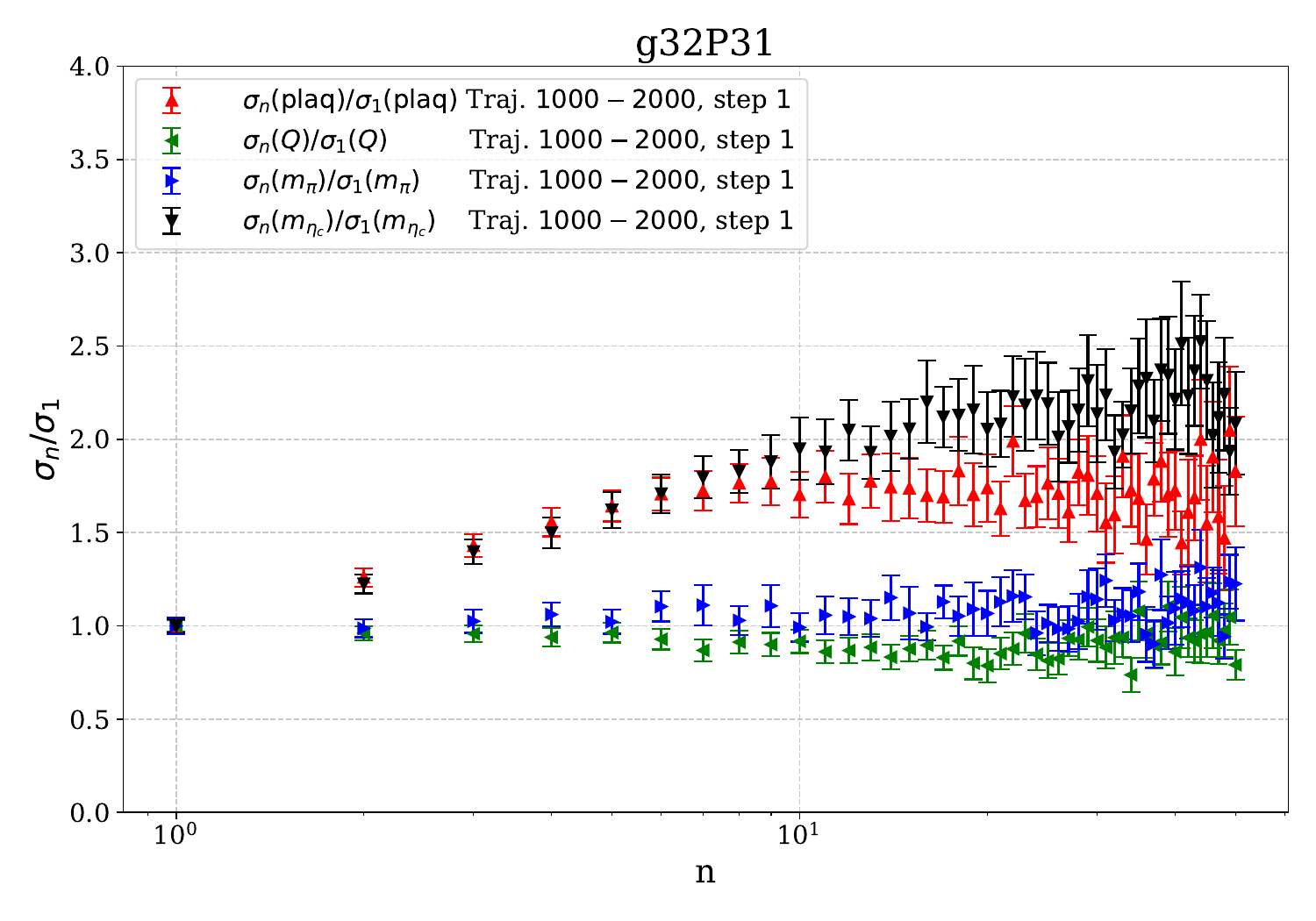} &
    \includegraphics[width=0.33\textwidth]{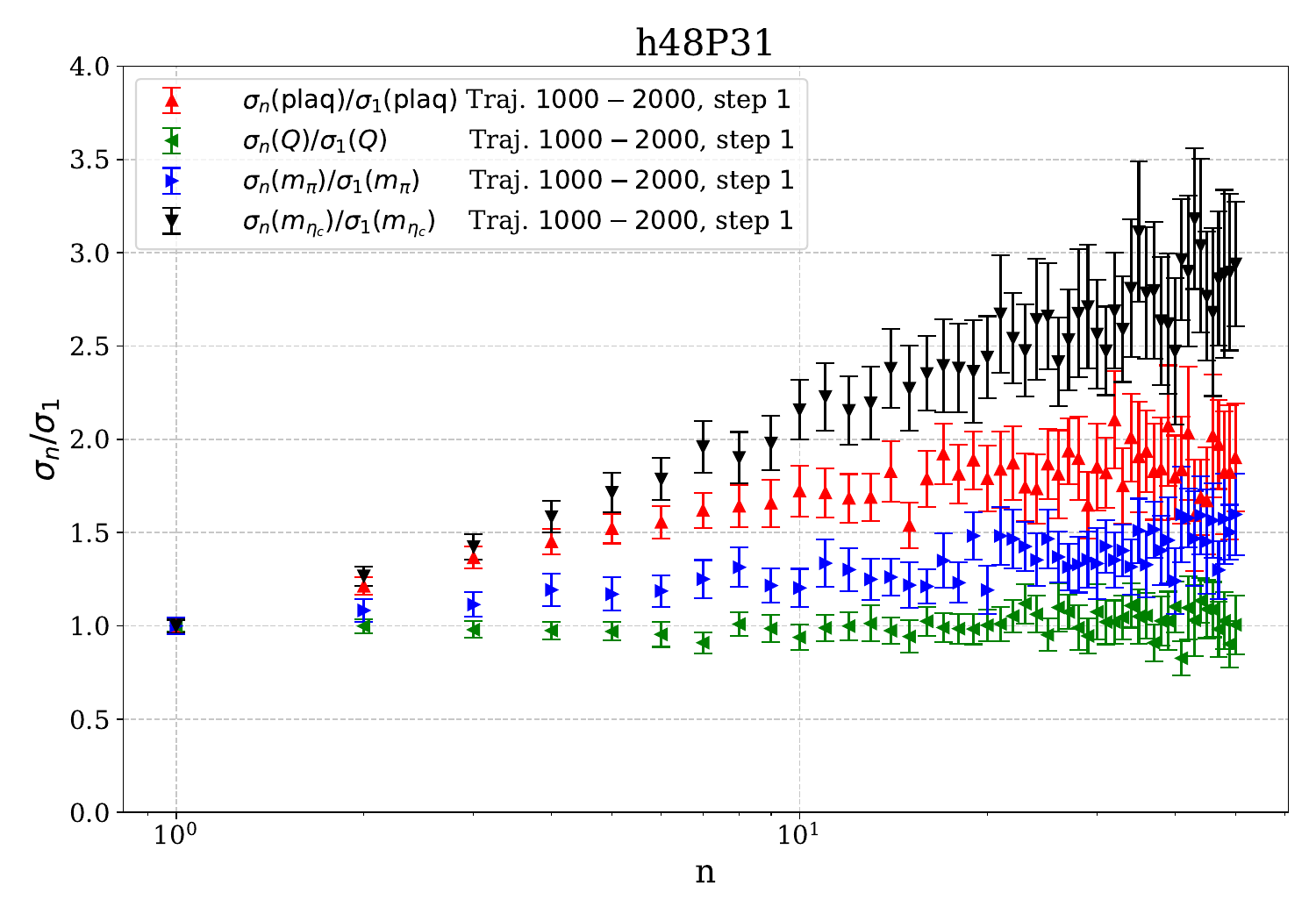} \\
    \includegraphics[width=0.33\textwidth]{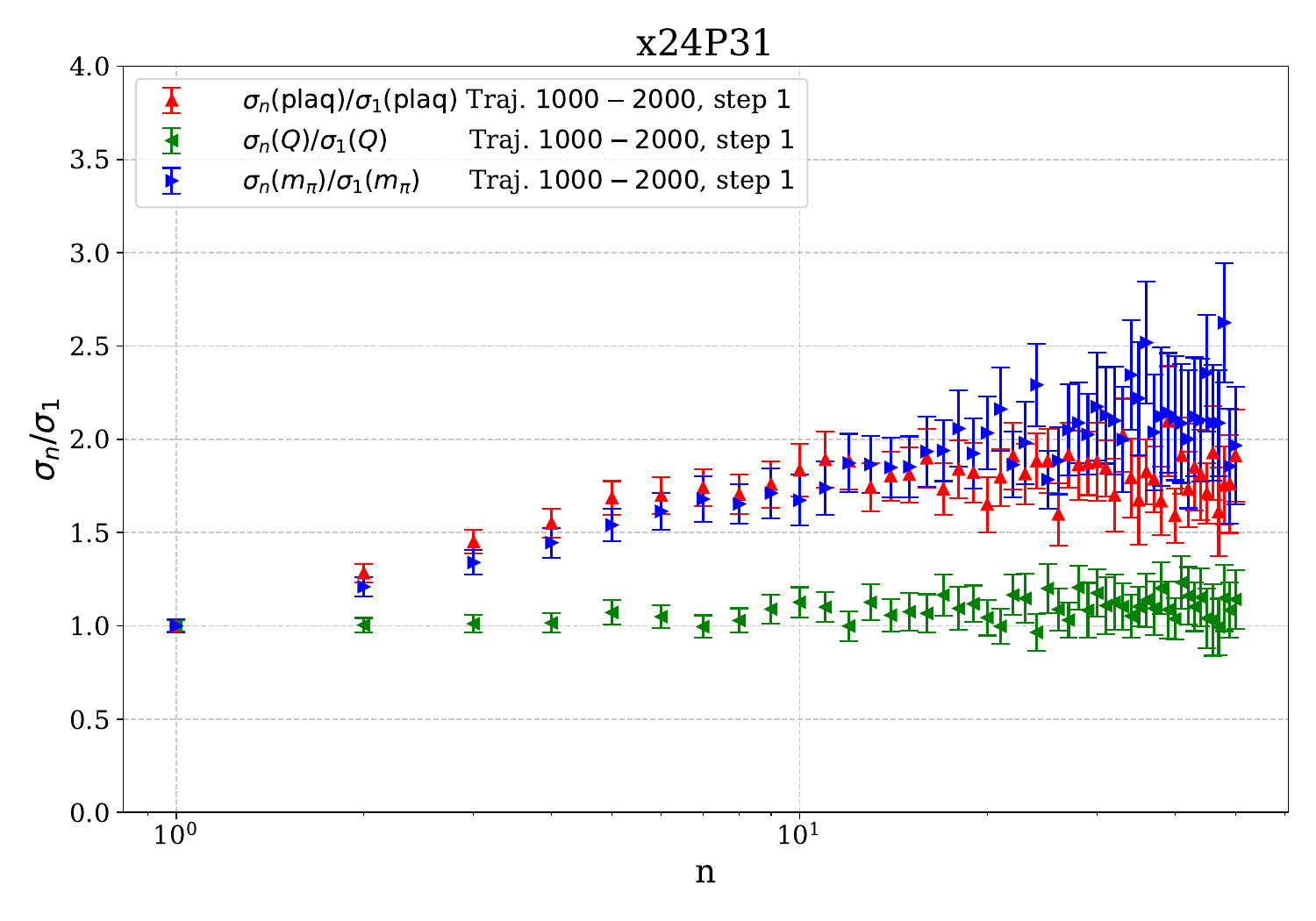} &
    \includegraphics[width=0.33\textwidth]{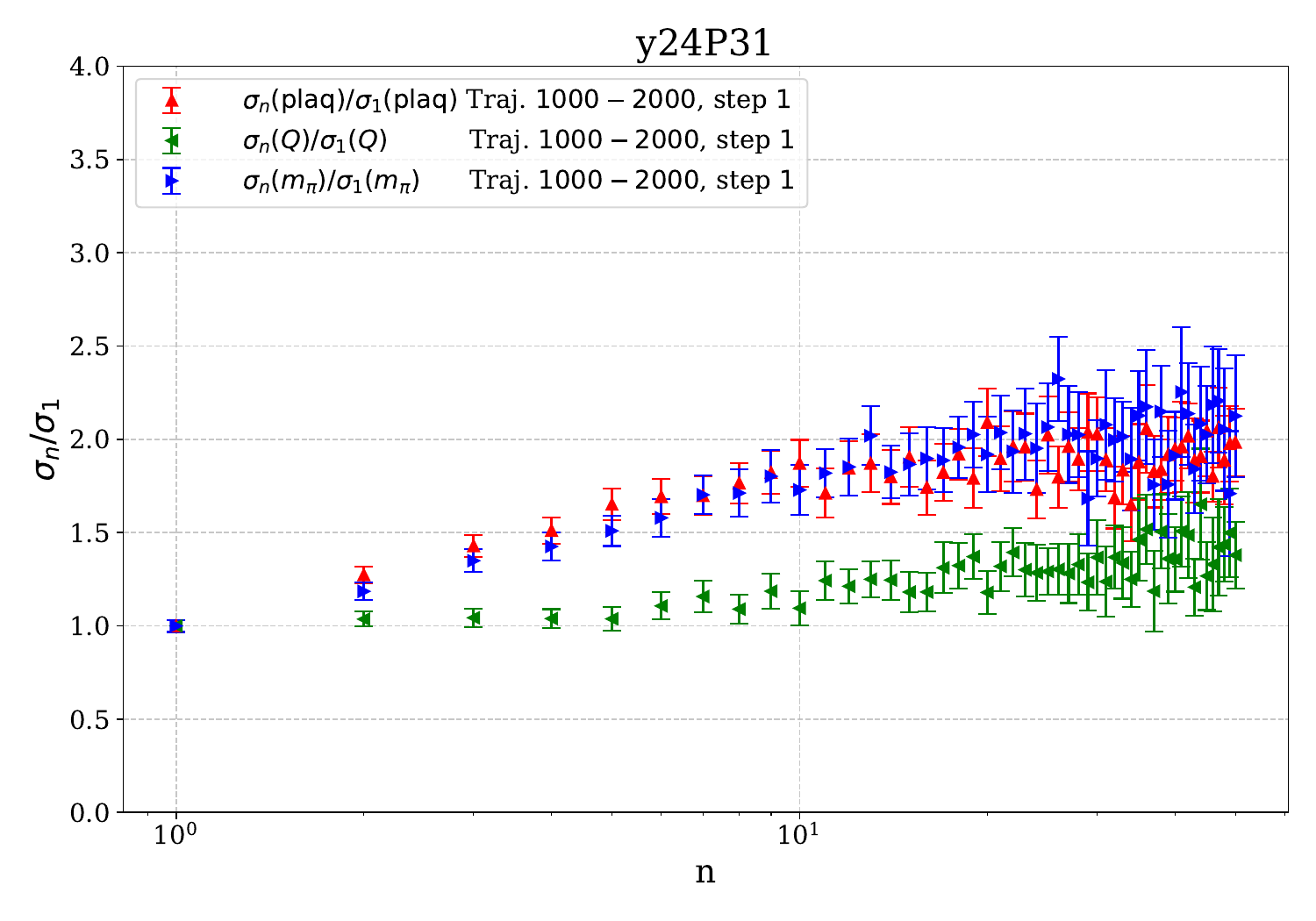} &
    \includegraphics[width=0.33\textwidth]{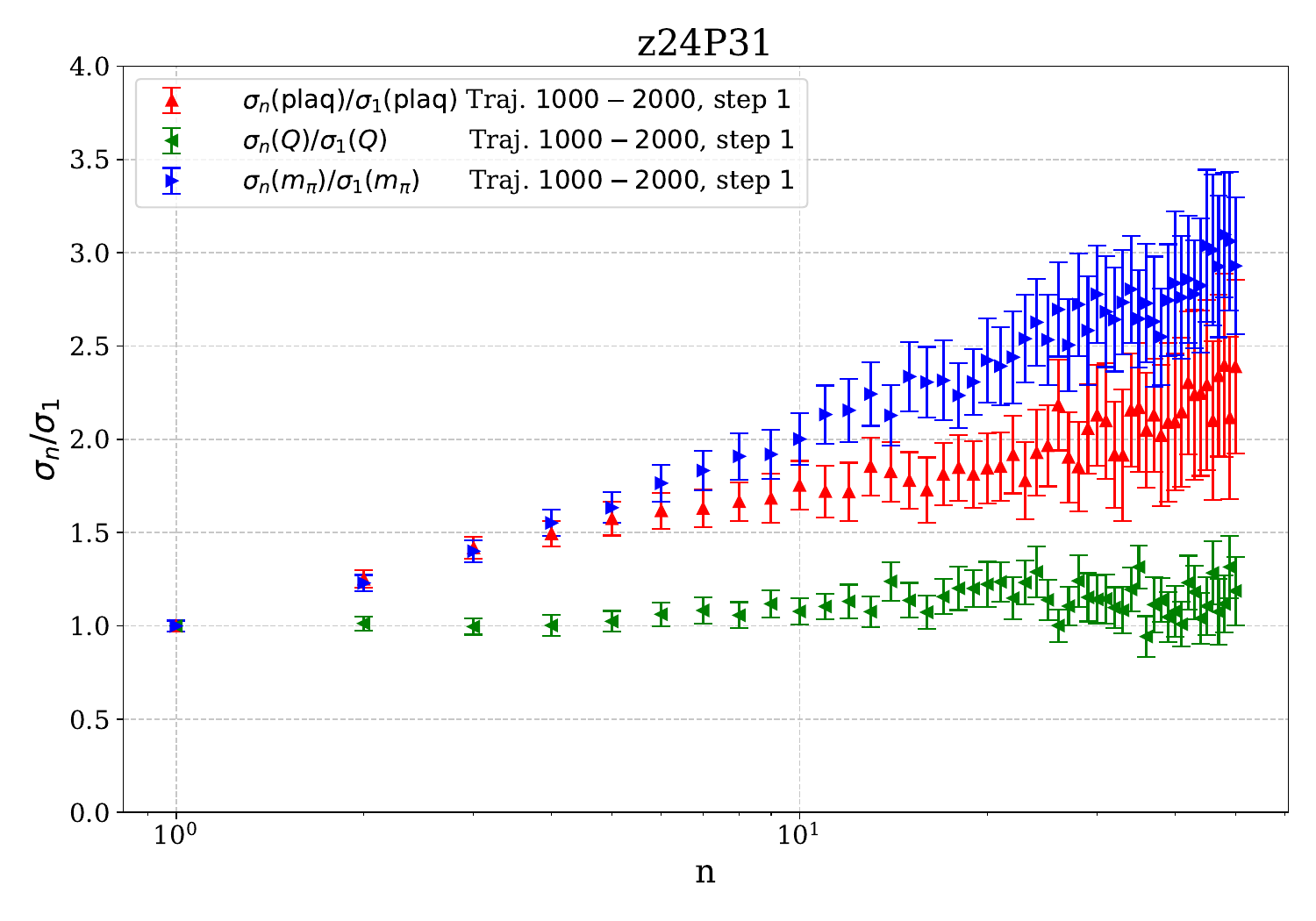} \\
    \end{tabular}
    \caption{Autocorrelation analysis across multiple lattice ensembles, depicting the bin-size dependence of variance for four fundamental quantities: the plaquette operator, topological charge, pion mass, and $\eta_c$ meson mass. The notation $\{\mathrm{Traj.\ }n_{\mathrm{min}},-n_{\mathrm{max}}, \mathrm{step\ }\Delta n\}$ accompanying each observable specifies the minimal and maximal trajectory indices, along with the stride parameter employed for configuration subsampling to ensure statistical independence in our analysis.}
    \label{fig:autocorr_composite}
\end{figure}

The upper three panels of Fig.~\ref{fig:autocorr_composite} illustrate the ratio $\sigma_n/\sigma_1$ as a function of bin size $n$ for all four observables, using data from the ensemble with $m_{\pi} \sim 0.3$ GeV at $a = 0.108$ fm. These panels correspond to different molecular dynamics time steps: $\tau = 0.5$ (upper left), $1.0$ (central left), and $2.0$ (lower left). As $\tau$ increases, the dependence of the statistical error on bin size weakens for all quantities. 

The three panels in the middle present the cases for ensembles at finer lattice spacings ($a = 0.09$, $0.07$, and $0.05$ fm from top to bottom), all with $\tau = 2$. These plots suggest that the sensitivity to binning does not become significantly stronger at smaller lattice spacings.

The lower three panels show the cases with different gauge actions with increasing $|C_P|$, tree level Symanzik, tadpole improved Symanzik, and Iwasaki. We can see that the autocorrelation effect is somehow stronger for the pion mass with the Iwasaki gauge action.

In summary, we observe weaker autocorrelation on the HISQ ensembles compared to the CLQCD ensembles, which we attribute to the larger Monte Carlo time ($\tau=2$) per trajectory in our simulations. These results ensure that the statistical uncertainties quoted throughout this work are not substantially affected by autocorrelation effects, thereby providing confidence in the reliability of our physical predictions.

\subsection{Interpolated parameters according to the lattice spacings of CLQCD ensembles}

Initial estimates for the parameters $\hat{\beta}$, $u_0$, and $\tilde{m}_s^{\rm b}$ at the lattice spacings of the $N_f=2+1$ CLQCD ensembles are obtained via second-order polynomial interpolation or extrapolation of our data; these values are collected in Table~\ref{tab:para_a}. The bare strange quark mass $\tilde{m}s^{\rm b}$ is rescaled to the value corresponding to the physical strange quark mass using ensembles with $m_s/m_l = 5$. It is then increased by 0.5\% to account for light quark mass dependence, based on the $m{\eta_s}$ values from the c24P31 and c48P13 ensembles.

These parameters can be further refined to obtain the same $L$ with specific spatial lengths $\tilde{L}=\{24,28,32,36,40,48,64\}$ by fine-tuning the lattice spacing; the resulting adjusted values are compiled in Table~\ref{tab:para_b}. It should be noted that these parameters serve as initial estimates and will require further adjustment during actual production to achieve self-consistent $u_0$, the target lattice spacing, and $m_{\eta_s}$ which corresponds to physcial strange quark mass.


\begin{table}[h] 
\centering
\caption{Initial guesses of $\hat{\beta}$, $u_0$, and $\tilde{m}_s^{\rm b}$ corresponding to the lattice spacings of the $N_f=2+1$ CLQCD ensembles.}\label{tab:para_a}
\begin{tabular}{ccccccc}
\hline \hline 
$a(\mathrm{CLQCD})$ &0.10542(64) &0.09015(58) &0.07761(46) &0.06896(44) &0.05239(32) &0.03761(23) \\
\hline
$\beta$ &7.3216(82) &7.4950(83) &7.6564(78) &7.7821(81) &8.0790(88) &8.531(32) \\
$u_0$ &0.88031(23) &0.88505(22) &0.88919(19) &0.89219(19) &0.89837(16) &0.90447(20) \\
$\tilde{m}^b_s$ &0.04590(38) &0.03838(34) &0.03222(28) &0.02798(26) &0.01988(20) &0.01268(21) \\
\hline \hline
\end{tabular}
\end{table}

\begin{table}[h] 
\centering
\caption{Initial guesses of $\hat{\beta}$, $u_0$, and $\tilde{m}_s^{\rm b}$ corresponding to the same $L\equiv \tilde{L}a$}\label{tab:para_b}
\begin{tabular}{ccccccc}
\hline \hline 
$a\,[\mathrm{fm}]$      &0.10417 &0.08929 &0.07812 &0.06944 &0.05208 &0.03906 \\
$\beta$ &7.3351(46) &7.5055(45) &7.6493(45) &7.7746(45) &8.0856(57) &8.462(21) \\
$u_0$   &0.88068(13) &0.88533(12) &0.88901(11) &0.89202(10) &0.89849(10) &0.90384(16) \\
$\tilde{m}^b_s$ &0.04528(21) &0.03796(18) &0.03248(16) &0.02822(14) &0.01973(13) &0.01339(17) \\
$\tilde{L}^3\times \tilde{T}$ & $24^3\times 72$ & $28^3\times 84$ & $32^3\times 96$ & $36^3\times 108$ &$48^3\times 144$ &$64^3\times 192$\\  
\hline \hline
\end{tabular}
\end{table}

\end{document}